\documentclass[fleqn,usenatbib]{mnras}

\newcommand{\flux}{\,erg\,s$^{-1}$\,cm$^{-2}$} %
\newcommand{\lumi}{\,erg\,s$^{-1}$} %
\usepackage[T1]{fontenc}
\usepackage{ae,aecompl}

\usepackage{graphicx}	
\usepackage{amsmath}	
\usepackage{amssymb}	
\usepackage[export]{adjustbox}
\usepackage[symbol]{footmisc}
\usepackage{fp}
\usepackage{multirow}
\usepackage{footnote}
\usepackage{microtype}

\title[K2-198 exoplanet evaporation]{Three young planets around the K-dwarf K2-198: High-energy environment, evaporation history and expected future}

\author[L. Ketzer et al.]{
L. Ketzer$^{1, 2}$\thanks{E-mail: ketzer.laura@gmail.com},
K. Poppenhaeger,$^{1, 2}$\thanks{E-mail: kpoppenhaeger@aip.de},
M. Baratella$^{1}$,
E. Ilin$^{1}$
\\
$^{1}$Leibniz Institute for Astrophysics Potsdam (AIP), An der Sternwarte 16, 14482 Potsdam, Germany\\
$^{2}$Universit\"at Potsdam,  Institut f\"ur Physik und Astronomie,  Karl-Liebknecht-Stra\ss e 24/25, 14476 Potsdam, Germany}

\date{Accepted 19/10/2023. Received 18/09/2923; in original form 15/05/2023.}

\pubyear{2023}

\begin{document}
\label{firstpage}
\pagerange{\pageref{firstpage}--\pageref{lastpage}}
\maketitle

\begin{abstract}
Planets orbiting young stars are thought to experience atmospheric evaporation as a result of the host stars' high magnetic activity. We study the evaporation history and expected future of the three known transiting exoplanets in the young multiplanet system K2-198. Based on spectroscopic and photometric measurements, we estimate an age of the K-dwarf host star between 200 and 500 Myr, and calculate the high-energy environment of these planets using eROSITA X-ray measurements. We find that the innermost planet K2-198c has likely lost its primordial envelope within the first few tens of Myr regardless of the age at which the star drops out of the saturated X-ray regime. For the two outer planets, a range of initial envelope mass fractions is possible, depending on the not-yet-measured planetary mass and the stars’ spin-down history. Regarding the future of the system, we find that the outermost planet K2-198b is stable against photoevaporation for a wide range of planetary masses, while the middle planet K2-198d is only able to retain an atmosphere for a mass range between $\sim$7 and 18\,$M_\oplus$. Lower-mass planets are too susceptible to mass loss, and a very thin present-day envelope for higher-mass planets is easily lost with the estimated mass-loss rates. Our results support the idea that all three planets started out above the radius valley in the (sub-)Neptune regime and were then transformed into their current states by atmospheric evaporation, but also stress the importance of measuring planetary masses for (young) multiplanet systems before conducting more detailed photoevaporation simulations.
\end{abstract}




\section{Introduction}

Young planetary systems ($\leq$~1\,Gyr) offer snapshots of the earlier stages of planetary evolution and can give insights into how planets have formed and are evolving. The first few hundred million years are expected to be most formative, with different physical processes shaping the planetary systems and their architectures. Processes that are thought to influence the evolution of young planetary systems are the shrinking and contraction of the cooling planet itself \citep[e.g.][]{2006Baraffe, 2012Mordasini}, orbital migration \citep[for a review, see][]{2014Baruteau}, planet-planet scattering \citep[for a review, see][]{2014Davies}, tidal and magnetic star-planet interaction \citep[e.g.][]{2011Poppenhaeger, 2017Strugarek, 2018Shkolnik, 2022Ilic}, as well as processes that drive the atmospheric mass loss of the primordial planetary atmosphere \citep[e.g.][]{2015Liu, Watson1981, 2018Ginzburg}.

The use of radial velocity follow-up and transit timing variations (TTVs) in measuring mass, as demonstrated in studies by e.g. \citet{2014Marcy, 2015Rogers, 2016JontofHutter}, has revealed that planets with radii smaller than approximately 2\,$R_\oplus$ have densities that suggest a rocky composition similar to Earth, while those with larger radii have lower densities indicating the presence of gaseous envelopes. These primordial atmospheres are thought to be mostly composed of hydrogen and helium and susceptible to mass loss on lower-mass planets. Giant impacts \citep[e.g.][]{2015Liu, 2020Wyatt} can explain some mass loss, but the two leading physical processes are photoevaporation, an externally induced mass loss mechanism where the star's high energy radiation (X-rays and extreme UVs, together XUV) heats the upper atmosphere and launches a hydrodynamic outflow \citep[e.g.][]{Watson1981, Guedel2007, 2012Owen_Jackson, Lopez2012}, and core-powered mass loss \citep[e.g.][]{2018Ginzburg, 2019Gupta}, which is driven by the internal luminosity given off by the cooling planetary core. Both processes are able to explain the prominent features in the observed exoplanet population: the radius gap, a deficit of planets with radii around 1.8\,$R_\oplus$, separating the rocky planet population below the gap from the planets with volatile, gaseous envelopes above \citep{Fulton2017, VanEylen2018b}, the hot Neptunian desert at very short orbital periods \citep{2016Mazeh, 2016Lundkvist}, and the eccentricity distribution of exoplanets \citep[e.g.][]{2020Correia}. Recently, gas accretion during the gas-poor phase of disk evolution \citep{2022Lee}, and the existence of water- and ice-worlds \citep[e.g.][]{2019Zeng, 2020Venturini} have also been put forth as mechanisms to explain the radius dichotomy of close-in exoplanets.

The gaseous envelopes surrounding Neptunes and mini-Neptunes are prone to escaping, leading to the observation of planetary tails comprised of escaping gas detected through the hydrogen Lyman-alpha line \citep[e.g][]{2015Ehrenreich, 2018Bourrier, 2022Zhang}, and more recently through helium absorption, as demonstrated in studies by e.g. \citet{2018Spake, 2018Mansfield, 2023Damasso}. Detecting outflows in young planetary systems is particularly interesting because the strength of the mass loss can  put constraints on planetary evolution models and might help to distinguish between different mass loss processes. While the discovery and subsequent characterization of young planets is challenging due to the high activity level of the host star, the number of young planets is increasing thanks to dedicated campaigns, giving insights on the role of photoevaporation in early exoplanet evolution. Interesting systems with detected outflow signatures are K2-100b, a $\sim$750\,Myr old, highly irradiated planet right at the border of the hot Neptunian desert with notable ongoing evaporation, causing the planet to significantly decrease its size over the next few Gyr \citep{2019Barragan, 2020Gaidos}. \citet{2022Zhang, 2023Zhang} further detect atmospheric escape for four young (<1\,Gyr) mini-Neptunes, predicting the loss of the remaining hydrogen-rich atmospheres for all the planets. These results indicate that for planets orbiting sun-like stars, photoevaporation is an efficient mechanism for stripping primordial gaseous atmospheres, transforming planets from mini-Neptunes above the radius gap into super-Earths below the gap.

In extremely young multiplanet systems like V1298 Tau and AU Mic ($\leq$~25\,Myr) \citep{2021Vissapragada, 2020Carolan, 2022Cohen}, as well as more mature systems like TOI-560 or K2-136 ($\sim$500-700\,Myr) \citep{2022Barragan, 2022Fernandez} all planets have experienced the same evolution of the high energy emission caused by the spin down and subsequent magnetic activity decay of the star. While this is not only unique to young systems, but also older ones, young multiplanet systems are of particular importance because the most extreme mass loss occurs early on. Depending on the properties of the planet and the host star, young planets can undergo drastic changes in size and mass within the first few 100 Myr. 
The photoevaporative mass loss of a planet can be quantified in terms of the mass-loss rate, an instantaneous quantity measuring the atmospheric mass loss at any given point in time, or the mass-loss timescale, the corresponding time-integrated quantity, and a measure of the total amount of envelope mass lost within a chosen time interval. 
The mass-loss timescale depends sensitively on the total amount of XUV exposure, which is influenced by the time at which the star has dropped out of the saturated X-ray regime, and two planets of similar age could have received an accumulated XUV flux that differs by a factor of 10 between a low activity star that drops out of the saturated regime after a few Myr, versus a highly active star that stays saturated for a few 100\,Myr (see e.g. Fig.~3 in \citet{2023Ketzer}). In multiplanet systems, we can conduct comparative mass-loss studies where the stellar contribution is controlled for. Since observable properties like the planetary radius depend strongly on the mass-loss history of the planet, well-constrained systems with planets below and above the radius gap in some cases can be used to constrain the rotational history of the host star \citep{2019Kubyshkina_a, 2019Kubyshkina_b, 2021Bonfanti}. \citet{2020Owen} also showed that systems with planets straddling the radius valley can be used to estimate the minimum mass of planets above the valley with remaining gaseous envelope. This makes multiplanet systems interesting targets for dedicated observing campaigns to measure masses as well as ongoing mass loss.

In this paper, we focus on characterizing the three-planet system K2-198, which was flagged as an interesting target for future atmospheric mass loss studies in \citet{2022Foster}. The host star, a K-dwarf, is orbited by three known transiting exoplanets discovered in the \emph{Kepler} K2 data \citep{2018Mayo, 2019Hedges}.
The innermost planet~c resides well below the radius gap in the regime of rocky planets, while the middle planet~d, with a radius of 2.4\,$R_\oplus$, sits just above the radius gap in the regime of mini-Neptunes. The outermost planet~b has a Neptune-like radius, placing it well above the radius rap in the regime of planets with significant volatile envelopes \citep{Otegi2020} (see Table~\ref{table:sys_parameters} for a summary of stellar and planetary parameters).
K2-198 was detected in X-rays by eROSITA with an X-ray luminosity of $7.9 \times 10^{28}$\,\lumi in the 0.2-2.0~keV band \citep{2022Foster}. Such X-ray luminosity suggests an active host star, placing the planets into an intense high-energy irradiation regime. 
The combination of an X-ray active host star, orbited by a close-in, likely rocky planet which has lost its primordial hydrogen-helium envelope, a planet straddling the top of the radius gap at a slightly larger orbit, and a  Neptune-sized planet further out, make this system of great interest for atmospheric mass loss studies. Further observations and studies of this relatively unexplored three-planet system may provide valuable insights into the formation and evolution of close-in planets and their atmospheres.

In this work, we use photometric and spectroscopic observations to estimate the age of the K2-198 system (Sec.~\ref{sec:age}), and characterize the current high energy environment (extreme UV and X-ray; together: XUV) and irradiation received by the three exoplanets in the system using updated eROSITA X-ray measurements. In Sec.~\ref{sec:atm_evo}, we re-estimate the current mass loss rates of the two outer planets and calculate their expected mass loss evolution over time scales of gigayears, using the code PLATYPOS\footnote{\url{https://github.com/lketzer/platypos/}}, which is based on an energy-limited evaporation model \citep{2022Ketzer}. We also explore the past of the planetary system to test the possible range of initial conditions, taking into account reasonable stellar spin-down and thus activity evolutionary tracks.

\begin{table}
\caption{Properties of the K2~198 star-planet system as provided by \citet{2018Mayo, 2019Hedges}.}
\centering
\label{table:sys_parameters}
\begin{tabular}{ll}
\hline \hline
Parameter & Value \\ \hline
\textit{Star:} & \\
Spectral type &  K-dwarf  \\
$M_\star$ [$M_\odot$] & 0.799$^{+0.045}_{-0.091}$  \\
$R_\star$ [$R_\odot$] & 0.757$^{+0.035}_{-0.016}$  \\
$T_\mathrm{eff}$ [K] & 5212.9$^{+49.2}_{-99.0}$  \\
Distance [pc] & 110.56$^{+0.87}_{-0.86}$  \\ [0.2cm]

\textit{planet~b:} & \\
$P$ [d] & 17.0428683$^{+0.0000035}_{-0.0000071}$  \\
$R_P$ [$R_\oplus$] & 4.189$^{+0.228}_{-0.098}$ \\
Semi-major axis ($a/R_\star$) & 25.86$^{+0.95}_{-0.48}$ \\ [0.1cm]

\textit{Planet c:} & \\
$P$ [d] & 3.3596055$^{+0.0000040}_{-0.0000021}$  \\
$R_P$ [$R_\oplus$] & 1.423$^{+0.081}_{-0.036}$ \\
Semi-major axis ($a/R_\star$) & 8.76$^{+0.32}_{-0.16}$ \\ [0.1cm]

\textit{planet~d:} & \\
$P$ [d] & 7.4500177$^{+0.0000026}_{-0.0000052}$  \\
$R_P$ [$R_\oplus$] & 2.438$^{+0.130}_{-0.056}$ \\
Semi-major axis ($a/R_\star$) & 14.90$^{+0.55}_{-0.28}$ \\

\hline
\end{tabular}
\end{table}


\section{Observations and data analysis}

The system was observed in X-rays by the eROSITA space telescope, as well as in the optical with TESS and K2. 

\subsection{X-ray data}

The position of the K2-198 system was observed by eROSITA \citep{Predehl2021}, an X-ray instrument onboard the Spectrum-R\"ontgen-Gamma spacecraft \citep{Sunyaev2021}. eROSITA is sensitive to photons in the energy band from 0.2-10~keV, and started an all-sky survey in 2019, where the whole sky is scanned every six months in great circles roughly perpendicular to the ecliptic. The position of K2-198 has been covered by five eROSITA All-Sky Surveys, called eRASS1 to eRASS5.

The X-ray luminosity of the host star K2-198 was first reported by \citet{Foster2022AandA} to be $7.9\times 10^{28}$\,erg/s in the 0.2-2~keV energy band, based on a detection of the star in the eRASS1 survey. Since then, data from the following four eRASS surveys (eRASS2 to eRASS5) has been made available to the eROSITA-DE consortium. The survey data sets were processed by the eRSOITA-DE consortium; specifically, a single-band detection in the 0.2-2.3~keV band was run on the stacked data from all five surveys, which was used as the input catalog for a forced photometry in the individual eRASS surveys\footnote{The following consortium catalog versions all\_eN\_SourceCat1B\_221031\_poscorr\_mpe\_photom.fits with N$=1...5$ were used.}.

We cross-matched the resulting catalog with the position of K2-198 within a $10^{\prime\prime}$ radius, and we checked that the matched X-ray sources are likely to be of stellar nature, along the lines of \citet{Foster2022AandA}. In this way, we found two detections and three non-detections for the five surveys available for K2-198's position. Specifically, K2-198 was detected individually in the eRASS1 and eRASS5 surveys, undetected in the individual eRASS2, 3 and 4 surveys, and detected in the stacked data from all five surveys together.

\subsection{K2 and TESS photometry}

K2-198 was observed twice with the \emph{Kepler} space telescope during the extended \emph{K2} mission in 2014 and 2018 (Campaign 6 and 17), and once by the transit exoplanet survey satellite (TESS) in 2021 (Sector 46). We use the Python package \emph{lightkurve} \citep{2018Lightkurve} to download the pre-processed lightcurves available for K2-198, which have been corrected for instrumental systematics related to the spacecraft using different detrending methods. We visually inspect the lightcurves for periodic amplitude modulations likely caused by star spots, and derive rotation periods using the Lomb-Scargle periodogram. We select the short and long cadence lightcurve products from the K2, EVEREST, and K2SFF pipeline \citep{2012Stumpe, 2014Vanderburg, 2018Luger}, as well as the 20-sec and 2-min cadence lightcurves from the TESS SPOC pipeline \citep{2016Jenkins}, normalize the lightcurves, measure the highest amplitude peak in each periodogram, and take the mean of all peak frequencies to determine the value for the rotation period of the exoplanet host star.

\subsection{TRES spectral analysis}

In order to constrain the age of K2-198, we derive the lithium (Li) and barium (Ba) abundances by synthesizing the 6708\,\AA\, and 5853\,\AA\, regions of the stellar spectrum. We refer the reader to Sect. 3.3.3 for a detailed discussion.

We downloaded the extracted TRES spectrum available on the ExoFOP website\footnote{\url{https://exofop.ipac.caltech.edu/tess/}}, with a resolution R=44000 and a SNR per resolution element of 52.9. In particular, we extracted and analyzed only order 38 (for Li) and 31 (for Ba), which contain the two spectral lines of interest. We also found an existing HIRES spectrum, which, however, is of such poor quality preventing us from a good Li or Ba detection.

We synthesized the two spectral regions using the code MOOG \citep{1973sneden} and creating the line lists with \texttt{linemake} \citep{2021placco}, which has the most up-to-date database of experimental atomic parameters of each spectral line. For the model atmosphere, we linearly interpolated from the ATLAS9 grid of \citet{2003castelli}, with solar-scaled chemical composition and new opacities (odfnew). We did not perform a complete abundance analysis, but we estimated the effective temperature ($T_{\mathrm{eff}}$) using magnitudes from the 2MASS \citep{2mass} and \textit{Gaia} DR3 \citep{2016gaia, gaiadr3} catalogs with the \texttt{colte} program by \cite{2021casagrande} (adopting the \textit{E(B-V)} value from the TIC catalog, \cite{2021paegert}). The $T_{\mathrm{eff}}$ spans values from 5190$\pm$62 in \textit{(R$_{p}$-K)} to 5273$\pm$68\,K in \textit{(G-J)}, and a weighted mean of 5225$\pm$40\,K. We estimated the surface gravity and the microturbulence following the same approach in \cite{2020baratella_ges}, finding 4.60$\pm$0.09\,dex and 0.82$\pm$0.05\,km\,$\rm{s}^{-1}$. For the synthesis, we adopted a $v \sin i$ values from 4 and 5\,km\,$\rm{s}^{-1}$ making sure that the profiles of other nearby lines match the observed spectrum.

\section{Results}

To characterize the high-energy environment of the multiplanet system and estimate the atmospheric mass loss of the three planets, we derive the current X-ray irradiation of the planets and constrain the age of the system using photometric and spectroscopic data. 

\subsection{eROSITA X-ray luminosity}
\label{subsec:Xrays}

The X-ray source matched with K2-198's position in the stacked eROSITA data, and has a nominal catalog flux of $3.8\times 10^{-14}$\flux\, in the 0.2-2.3~keV energy band, representing the average flux of the star. However, the nominal catalog fluxes were calculated assuming an underlying power law, while stellar coronae have an underlying optically thin thermal plasma spectrum. Following the analysis of \citet{Foster2022AandA}, we therefore multiply the catalog fluxes by a conversion factor of 0.85 to derive coronal fluxes of the star in the more commonly used 0.2-2~keV energy band. 

In this way, we derive stellar X-ray fluxes of $3.2\times 10^{-14}$\flux\, (0.2-2~keV) as the average flux over the five eROSITA surveys, and $5.4\times 10^{-14}$\flux\, and $7.0\times 10^{-14}$\flux\, (0.2-2~keV) for the two individual detections in the eRASS1 and eRASS5 surveys. The non-detections in eRASS2, 3 and 4, which happen to have a shorter exposure time at K2-198's position, amount to X-ray flux upper limits of $8.5\times 10^{-14}$, $1.0\times 10^{-13}$, and $1.2\times 10^{-13}$\flux, using the same conversion factor.\footnote{The eROSITA upper limits (Tub{\'i}n-Arenas et al.\ 2023, submitted) were computed based on X-ray photometry on the eROSITA standard calibration data products (counts image, background image, and exposure time) and following the Bayesian approach described by \citet{KraftBurrowsNousek1991}. The upper limits are given as one-sided $3\sigma$ confidence intervals (99.87\%) and use the photons from eROSITA's 0.2-2.3~keV energy band.}

In our further analysis, we use the average X-ray flux over the five surveys, which is a lower value than the one reported by \citet{Foster2022AandA}, who only used the eRASS1 data, which apparently observed the star in a higher magnetic activity state. We note that since the number of excess counts is only on the order of 10 in surveys 1 and 5 together, a detailed spectral analysis with a fit of the coronal temperature is not feasible.

We use the \textit{Gaia} DR3 distance of 109.6\,pc to convert the measured average X-ray flux into an average X-ray luminosity of $L_X = 4.6 \times 10^{28}$ \lumi ($\log L_X$\,[\lumi]$\,=\,$28.7) \citep{2016gaia,2022Gaia}.

\subsection{Stellar rotation period and flaring activity}
\label{subsec:rotation_flaring}

All inspected light curves, covering a timespan from 2014 to 2021, show highly visible periodic brightness variations on the order of 2-3\%, typical for a young and active star with large star spots on the stellar surface \citet[e.g.][]{2016Stauffer}. From the Lomb-Scargle periodogram, we derive a mean rotation rate of $7.1 \pm 0.1$~days.

We used AltaiPony~\citep{Ilin2021} to inject synthetic flares into the Sector~47 TESS light curve, and quantify their recovery efficiency. Flares with energies below $\sim 10^{34}$\,erg are typically not recovered, so there could be at least one flare with $\sim 10^{34}$\,erg in the light curve. This gives a minimum flare rate of 15 flares per year above $\sim 10^{34}$\,erg in K2-198, which is up to an order of magnitude above the flare rate of stars with $T_\mathrm{eff} > 5000$\,K in the Pleiades~\citep{2021Ilin}, and in the range of the most active Sun-like stars~\citep{2013Shibayama}.

\subsection{Lithium and barium abundance}

From our analysis of the TRES spectrum, we derive a lithium (Li) abundance of A(Li)$_{\mathrm{LTE}}$= 2.07$\pm$0.08$\pm$0.13, where the first uncertainty is due to the fitting procedure and the second is the contribution of the atmospheric parameter uncertainties. With the NLTE corrections from \cite{2009lind}, we find a lithium abundance of A(Li)$_{\mathrm{NLTE}}$=2.15$\pm$0.08$\pm$0.13. In addition, we also derive a barium (Ba) abundance ratio of [Ba/H]=+0.41$\pm$0.10$\pm$0.09\,dex over the solar values from \cite{2021asplund}. The uncertainties are computed in the same way as for Li.

\subsection{Stellar age determination}
\label{sec:age}

To put the K2-198 system into context, and describe the high-energy irradiation environment of the three planets together with their past and future atmospheric evolution (see Sec.~\ref{sec:atm_evo}), we first need to constrain the stellar age of the system. We use the determined rotation period, X-ray activity level as well as independent spectroscopic age indicators to estimate an age between 200 and 500\,Myr, and adopt the logarithmic mean of 316\,Myr as the present age of the system.

\subsubsection{Rotation-based age}

\begin{figure}
\includegraphics[width=0.49\textwidth]{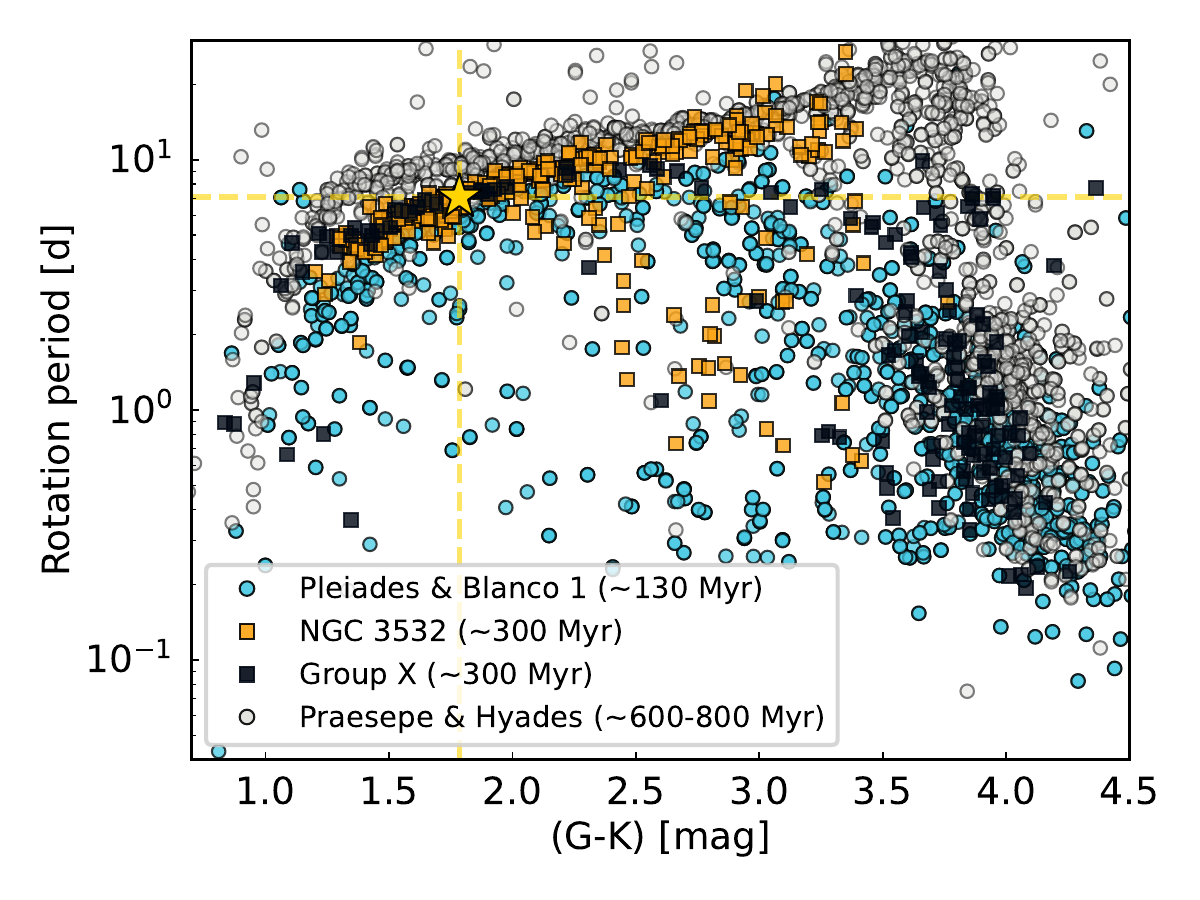}
\caption{Color-period diagram showing the rotational sequences for selected clusters of different ages. The location of K2-198, which is marked with a yellow star, indicates an age older than roughly 100\,Myr, but significantly younger than 600-800\,Myr.}
\label{fig:col_per}
\end{figure}

We use the periodic brightness variations from star spots as a proxy for the stellar rotation period, and by applying gyrochronology, to estimate an age of the system. 
For a K-dwarf, this well-constrained rotation period of roughly 7 days is still relatively short, indicating a young age (see e.g. \citet{Barnes2003}). Figure~\ref{fig:col_per} shows a color-period diagram for stars in the Pleiades, Blanco 1, NGC 3532, Group X, Praesepe and Hyades cluster, covering an age range from approximately 100 to 800\,Myr \citep{2020Curtis, 2020Gillen, 2021Fritzewski, 2022Messina, Wright2011, 2016Newton, 2019Douglas}\citep{2019Douglas, Wright2011}. By visual inspection, the K-dwarf K2-198 is located above the gyrochronological sequence of the Pleiades and Blanco~1 ($\sim$130\,Myr), and well below that of the Praesepe and Hyades ($\sim$600-800\,Myr). With its 7-day rotation period, the star can be placed nicely on the rotational sequence for the $\sim$300~$\pm$~60\,Myr old clusters NGC~3532 and Group~X, indicating the youth of the star-planet system. The absence of flares in the TESS light curve, as discussed in Sec.\,\ref{subsec:rotation_flaring}, is compatible with the estimated young age of K2-198.

We also make use of \textit{stardate} \citep{2019Angus}, a Python tool, which combines isochrone fitting with gyrochronology, to infer a Bayesian age for our system. The tool employs the affine invariant ensemble sampler \textit{emcee} \citep{2013ForemanMackey}. We use 50 walkers, 500,000 samples and a burn-in phase of 500, and obtain an age of $0.46_{-0.13}^{+0.04}$\,Gyr, in agreement with the location on the color-period diagram. 

\subsubsection{X-ray activity}

\begin{figure}
\includegraphics[width=0.492\textwidth]{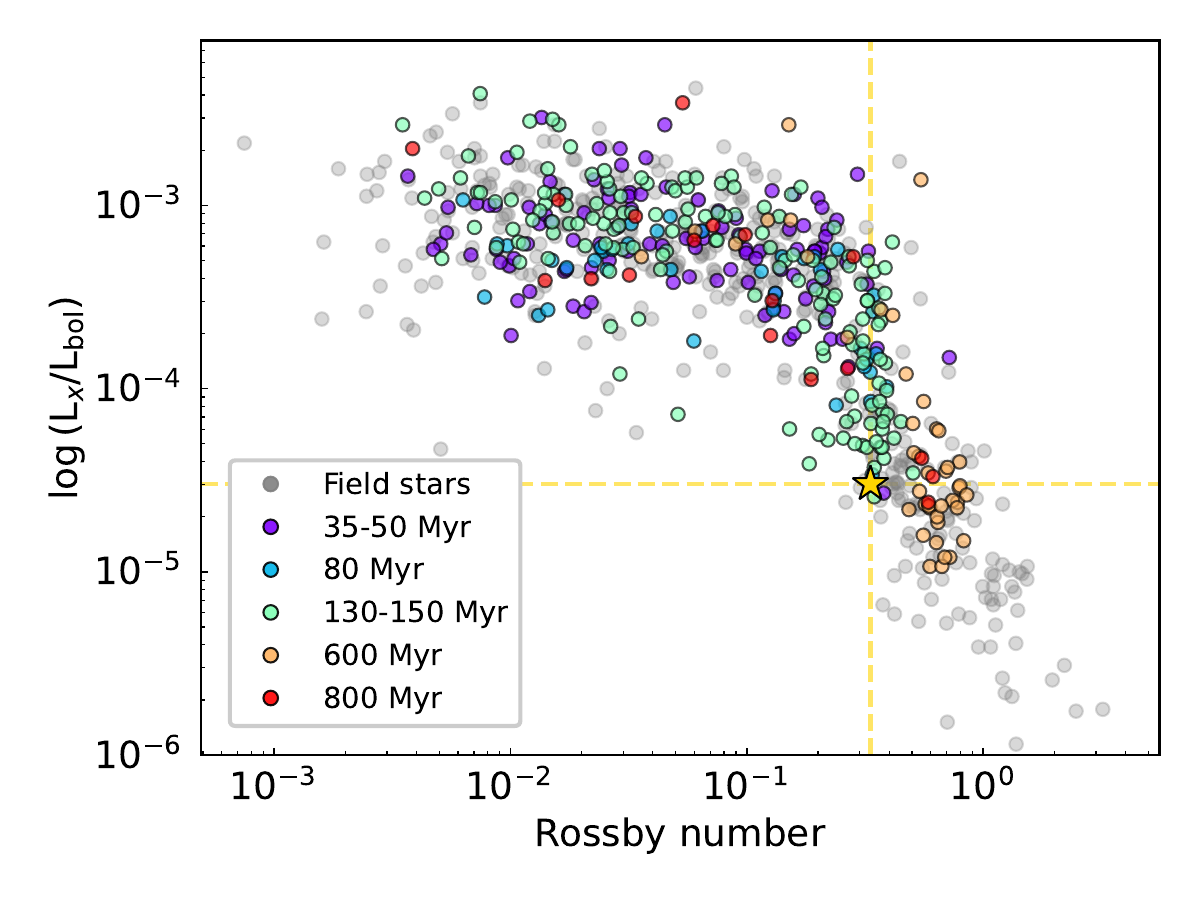}
\caption{Plot of $L_X/L_\mathrm{bol}$ as a function of Rossby number for all stars in the sample by \citet{Wright2011} with measured rotation periods. Field stars are marked in gray, while cluster stars with ages below $\sim$1\,Gyr are shown in five different-colored age bins. The location of K2-198, which is marked with a yellow star, indicates that the star has already dropped out of the saturated regime and concurs with an age between $\sim$~150 and 600\,Myr.}
\label{fig:Xray_Ro}
\end{figure}

In Figure~\ref{fig:Xray_Ro}, we show the X-ray luminosity normalized by the bolometric luminosity, $L_\mathrm{bol}$, as a function of the Rossby number for late-type stars in young open clusters of different ages. The data is taken from \citep{Wright2011}, and the conversion from stellar rotation period, P$_\mathrm{rot}$, to Rossby number, $Ro$, is done using the empirically determined convective turnover times, $\tau_{c}$, by \citet{2018Wright} (their Eq.~6) and is given by $Ro$ = P$_\mathrm{rot}$/$\tau_{c}$. The location of K2-198, with $L_{X}/L_\mathrm{bol} = 2.9\times10^{-5}$ and $Ro = 0.33$, indicates that the star is well beyond the breakpoint between the saturated regime, where $L_{X}/L_\mathrm{bol}$ is approximately constant, and the unsaturated regime, which stars enter during their spin-down phase as they age. A comparison with clusters of various ages points towards an age of K2-198 between $\sim$~150 and 600\,Myr, in agreement with the age determination from gyrochronology.

The X-ray fluxes reported in Sec.\,\ref{subsec:Xrays}  indicate that the star displays some variability over the observed timescale of roughly 2 years. Given the determined average X-ray flux and the larger value of the two individual detections, we estimate a variability amplitude of around 2.2. $\epsilon$\,Eridani ($\sim$~400 Myr), and $\iota$\,Horologii ($\sim$~600 Myr), the two youngest solar-like stars with detected X-ray activity cycles, exhibit not only the shortest X-ray cycles (with approximately 1.6 and 2.9 years), but also the smallest variations in X-ray luminosity (on the order of 2 throughout the cycle) among all stars with detected coronal cycles \citep{2019SanzForcada, 2020Coffaro}. Kepler-63, a fast rotator ($P_\mathrm{rot} \approx$~5.4 days) \citep{2013SanchisOjeda} with an estimated age of 200 Myr, did not display cyclic X-ray variability, suggesting that stars much younger than 400 million years might have inhibited X-ray cycles due to a significant presence of coronal magnetic regions \citep{2022Coffaro}. Although no cycle was detected, the X-ray data for Kepler-63, which covers just over a year, shows a minimum-to-maximum variation smaller than a factor 2. While the eROSITA data for K2-198 does not allow for detailed statements about cyclic variability, the data is compatible with the variability observed in other young stars.

\subsubsection{Lithium and barium}

It is well known that the photospheric abundance of Li decreases with increasing age in late-type stars. When a star is born, its Li abundance reflects the abundance of the interstellar medium from which it has formed. Then, thanks to several transport mechanisms, some of the photospheric Li is brought into deeper layers where it is exposed to temperatures larger than $\sim 2.6\times10^6$\,K, with the consequence of easily being destroyed. Therefore, Li is depleted in the photosphere as the star evolves by a factor 30-60 after the first dredge-up, and it can be used as powerful age diagnostics \citep{1967iben,2013jeffries,2021romano}.

In addition to Li, it has recently been demonstrated that in young stellar clusters (open clusters, moving groups and local associations), the Ba abundance increases dramatically with decreasing age, with values around [Ba/Fe]$\approx$+0.65\,dex at 50\,Myr \citep{2009dorazi,2015reddy,2018magrini}.
The same trend has been also observed in solar twins \citep{2017reddy}, where the authors, for the first time, showed an interesting correlation with activity. In \cite{2021baratella} it was finally demonstrated how the anomalous over-abundance does not result from peculiar nucleosynthesis, but is mostly related to alterations of the spectral line formation due to the more intense stellar activity at such young ages. This behavior is valid for all young/active stars, not only in open clusters, but also in the field (see also \citet{2022dorazi} for a complete review on the topic).
While the main process behind such alterations is not well understood yet, such chemical peculiarities can nevertheless be used to probe the youth of a star.

\begin{figure}
\includegraphics[width=0.49\textwidth]{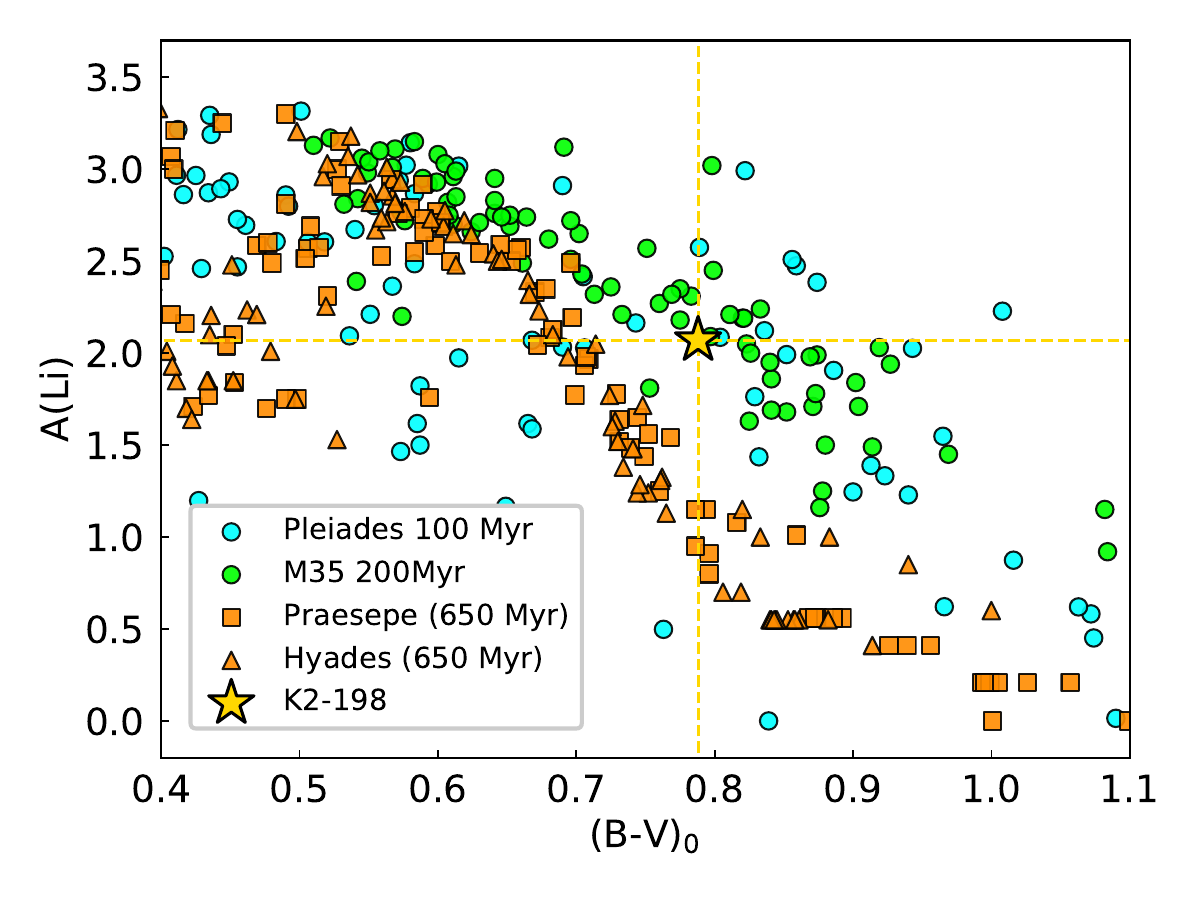}
\caption{Abundance of lithium as a function of the color index \textit{(B-V)$_{\mathrm{0}}$} of different young clusters. The stellar ages range from as young as 100-200\,Myr (circles) to roughly 650\,Myr, or the age of Praesepe and Hyades (squares and triangles). K2-198 is marked with a yellow star. For a K-dwarf, the measured lithium abundance indicates an age significantly younger than the Praesepe/Hyades.}
\label{fig:Li}
\end{figure}

In Figure \ref{fig:Li}, we plot the lithium abundance, A(Li), as a function of \textit{(B-V)$_{\mathrm{0}}$} for different young stellar clusters. We show the Pleiades ($\sim$100\,Myr, \citealt{2018bouvier}), M35 ($\sim$200\,Myr, \citealt{2018anthony}), the Hyades and Praesepe ($\sim$600-650\,Myr, \citealt{2017cummings}). Our target places near the Pleiades and M35 distribution, suggesting an age of $\sim$200\,Myr and definitely younger than the Hyades/Praesepe.
This is also corroborated by the Ba abundance, for which we found an abundance ratio of [Ba/H]=+0.41$\pm$0.10$\pm$0.09\,dex over the solar values from \cite{2021asplund}. The super-solar Ba abundance is similar to what is found at similar ages in Galactic open clusters in \cite{2021baratella} and \cite{2022dorazi}, suggesting an age significantly younger than 1\,Gyr.

\subsubsection{Kinematic Age Determination}

Another independent way to statistically estimate stellar ages is through the empirical age-velocity dispersion relation (short: AVR) \citep{1946Stromberg, 2009Holmberg}, which is based on the observation that older stars have a larger velocity dispersion. As part of their Planets Across Space and Time (PAST) project, \citet{2021Chen} characterize the membership of galactic components based on stellar kinematics (i.e., astrometry and radial velocities) provided mainly by Gaia and LAMOST, and derive stellar kinematic ages, with a typical uncertainty of 10-20\%, for a sample of 2000 exoplanet host stars. In their published catalog of kinematic properties, they report relative probabilities for stars belonging to different galactic components (e.g., thin/thick disk), which they show is correlated with stellar age (see their Fig.~18). The kinematic age generally increases with the relative membership probability, TD/D (thick disk over thin disk), illustrating that TD/D is indeed an indicator of age for stars in the Galactic disk. Fig.~18 in \citet{2021Chen} shows that the predicted age of K2-198 is well below 1\,Gyr, highlighting the youth of the system. With a reported relative probability, TD/D, of 1.85e-02, the predicted kinematic age is $\sim$0.84\,Gyr. Kinematic methods like the AVR are ideally used for an ensemble of stars and not for individual systems, which is why we do not include this age value in our final estimate. Nevertheless, it supports the notion that K2-198 is a young star, with age <1\,Gyr.

Based on the age constrained by the rotation period, X-ray activity and lithium abundance, we are confident about the youth of the system and estimate K2-198's age to lie between 200 and 500\,Myr. We adopt the logarithmic mean of 316\,Myr as the present age of the system, and use this as input for the atmospheric evolution and escape calculations.

\subsection{Atmospheric evolution and escape}
\label{sec:atm_evo}

We estimate the current atmospheric mass loss rate of the K2-198's two outer planets, which reside above the observed radius gap and likely still host gaseous envelopes. We perform two types of calculations: in Sec.~\ref{subsec:future} we estimate the future radius evolution of the two outermost planets to test whether one of the two planets can survive above the gap; and in Sec.~\ref{subsec:past}, we investigate the past of all three planets by calculating backwards in time.

All simulations are conducted with the publicly available python code PLATYPOS\footnote{\url{https://github.com/lketzer/platypos/}}. For a detailed description of the code and its limitations, see \citet{2022Ketzer} and Sec.~2 in \citet{2023Ketzer}. The code uses the formalism of energy-limited atmospheric escape, with the inclusion of a radiation/recombination-limit, to calculate atmospheric mass loss rates of planets with gaseous hydrogen-helium envelopes atop rocky cores. Unless stated otherwise, we adopt a heating efficiency of 0.1, calculate the XUV absorption radius according to \citet{2017Lopez}, and the extreme ultraviolet (EUV) flux based on the updated X-ray-EUV surface flux relation by \citet{2021Johnstone}. The planetary radius evolution is modelled using the mass-radius-age fitting formula by \citet{2016ChenRogers}, which takes into account that planets contract as they cool.

\begin{figure}
\includegraphics[width=0.5\textwidth,trim=0.9cm .0cm .0cm .0cm,clip]{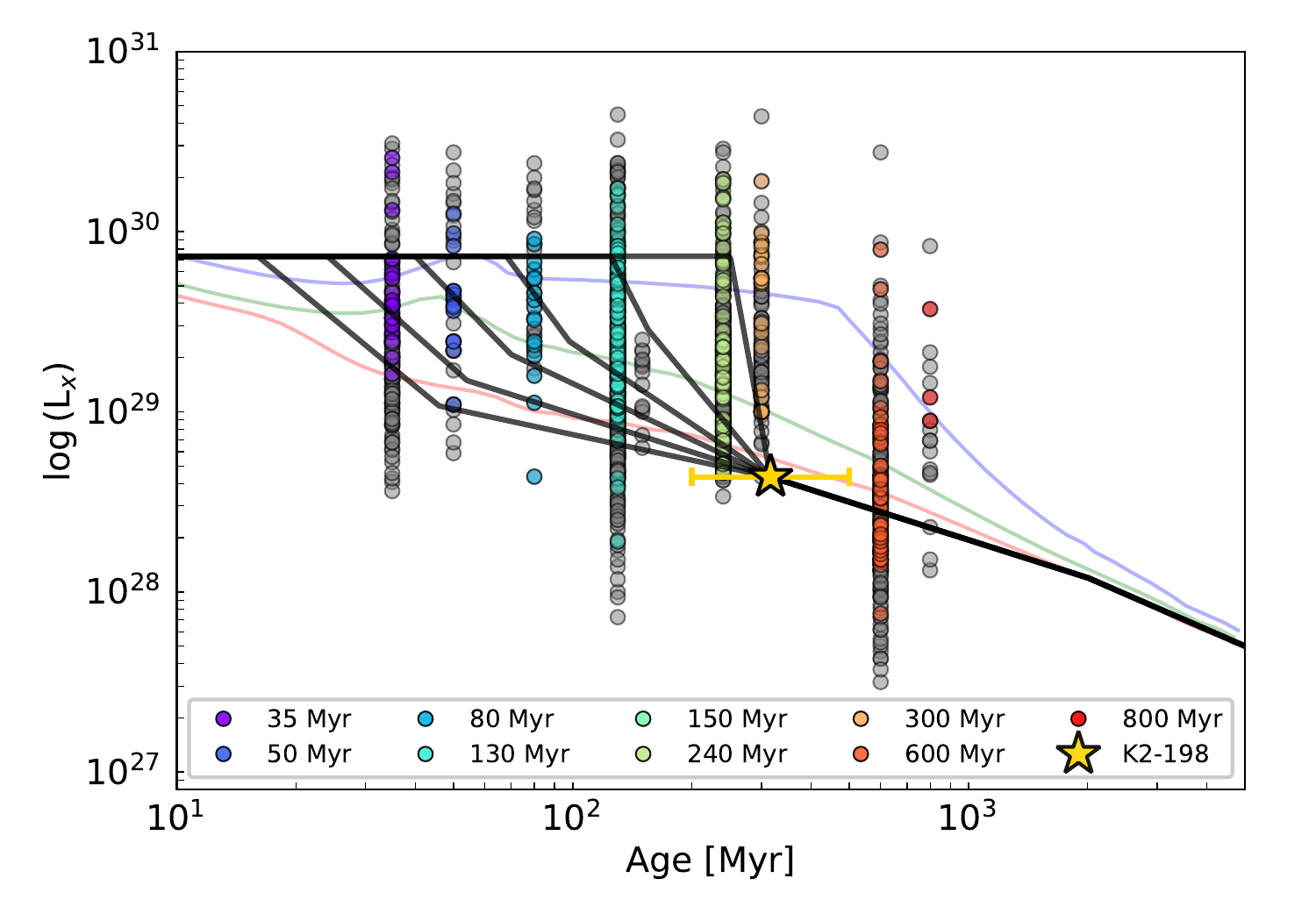}
\caption{Plot of $L_{X}$ as a function of stellar age for stars in open clusters below 1\,Gyr \citep{Wright2011}. We show all stellar types in gray, and color-code only the stars with stellar masses between 0.7 and 0.9\,M$_\odot$ by age. The translucent red, green and blue lines show stellar spin-down models for an $\sim0.8$\,M$_\odot$ star for an initially slow, intermediate and fast rotator \citep{2021Johnstone}. We plot K2-198 based on the measured eROSITA X-ray luminosity and the adopted age of 316\,Myr. The error bars mark our constrained age range of 200-500\,Myr. K2-198's location indicates that the star is on the inactive side for its age and spectral type, and has already dropped out of the saturated regime, in agreement with Fig.~\ref{fig:Xray_Ro}.}
\label{fig:Xray_age}
\end{figure}

Regarding the future activity evolution of the host star, based on the estimated age range of 200-500\,Myr and the measured X-ray luminosity of $4.6 \times 10^{28}$ \lumi, we infer that the star has already dropped out of the saturated regime, converged onto the slow-rotator sequence and will continue its activity decay along one specific track. For our investigation of the past of the star-planet system, we take into account that stars of similar spectral type drop out of the saturated regime over a range of ages, depending on initial rotation rate and/or magnetic field complexity \citep{2011Wright, 2015Matt, 2015Tu, 2018Gondoin, 2018Garraffo}. The stellar high-energy activity tracks used in this work are motivated by the spread in X-ray luminosities of young cluster stars of similar age (see Fig.\ref{fig:Xray_age}) \citep[e.g.][]{2011Wright}, and to some extent by the rotational spin-down, and thus activity evolution models by \citet{2021Johnstone}. We choose several ages, at which the star might have dropped out of the saturated regime, covering a range from low to high activity expected for K-dwarfs. The X-ray luminosity in the saturated regime is calculated according to the updated ${\log(L_{\mathrm{X}_{\mathrm{sat}}}/L_{\mathrm{bol}})}$-fit by \citet{2021Johnstone}.

\subsubsection{Planetary mass estimates and current mass loss rate estimates}
\label{sec:masses}

Since no planetary masses have been measured for this system, we use the observed radii and existing mass-radius relations to estimate masses for the three planets. For the innermost planet c, we assume it to be fully rocky, which, given the mass-radius relation for Earth-like rocky cores \citep{LopezFortney2014, 2016ChenRogers}, corresponds to a bare core of 3.9\,$M_\oplus$. The two outer planets d and b have radii which place them in the volatile regime according to \citet{Otegi2020}. Using their observationally-based mass-radius relation (see their Fig.~2), we estimate a realistic mass range of $3.6-22.0$\,$M_\oplus$ for planet~d, and $7.7-25.0$\,$M_\oplus$ for planet c. While the 2$\sigma$ envelope of the mass-radius relation by \citet{Otegi2020} predicts a mass as high as 60\,$M_\oplus$ for a planet with a radius of 4.1\,$R_\oplus$, our upper mass limit for the outermost planet is set by the grid-limits behind the mass-radius-age fitting formula by \citet{2016ChenRogers}. For the calculation of the planets' past in Sec.~\ref{subsec:past}, we restrict ourselves to three example masses, covering the estimated mass ranges for planets d and b.

At present, the X-ray irradiation levels of the three planets are $1.7\times10^{4}$, $6.0\times10^{3}$, and $2.0\times10^{3}$ \flux, going from the innermost planet c to the outer ones d and b. By estimating the EUV emission according to \citet{2021Johnstone}, we obtain XUV fluxes of $4.1$, $1.4$ and $4.7\times10^{3}$ \flux for planets c, d and b, respectively. Planet c resides below the exoplanet radius gap and is thus assumed to be an evaporated, leftover, bare rocky core. According to \citet{Fossati2017}, photoevaporation of planetary atmospheres occurs when the restricted Jeans escape parameter is smaller than 80, which is true for all the planets under consideration. Thus, assuming the two outer planets, which are located above the radius gap, still host volatile envelopes, we estimate present-day mass loss rates of $2.5\times 10^{9}$ to $3.7\times10^{10}$ g\,s$^{-1}$ for planet~d and $4.0\times10^{9}$ to $2.0\times10^{10}$ g\,s$^{-1}$  for planet~b, using the modelling assumptions and planetary masses described above.

\subsubsection{Future of the planetary system}
\label{subsec:future}

We evolve the two outermost planets d and b forward in time to investigate their future atmospheric mass loss and subsequent radius evolution, and to predict whether they will lose their remaining envelope and fall below the gap or not. Planet c already resides below the gap and thus is assumed to have lost its envelope by the present age (which we take to be 316\,Myr in the simulation). We terminate the simulations at a final age of 5\,Gyr, which is approximately solar system age and close to the median age of the observed exoplanet population \citep{2023Gaidos}, or the age at which the planet turns into a bare core.

\begin{figure*}
\includegraphics[width=0.98\textwidth]{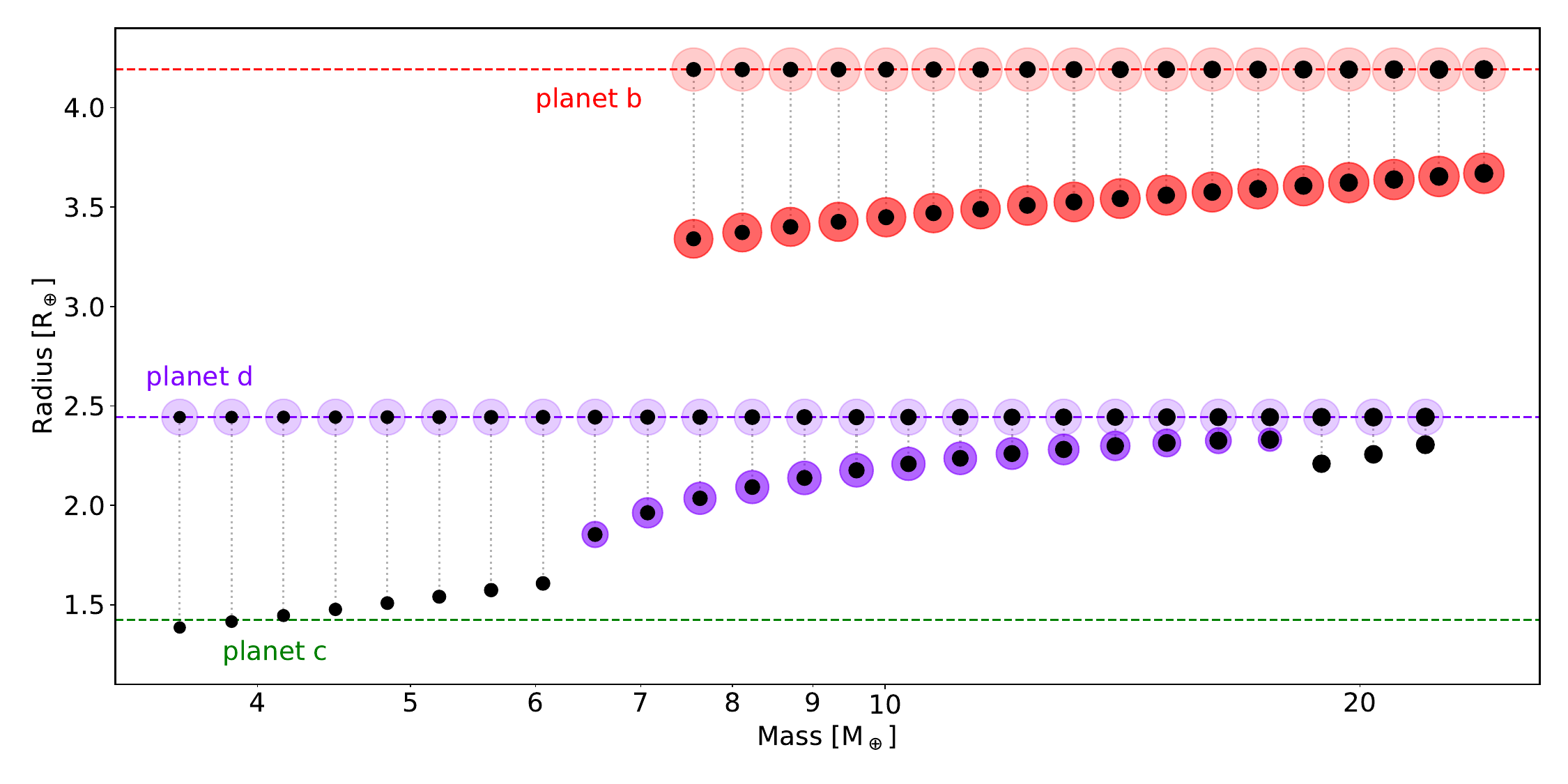}
\caption{Future radii of planets b, c, d at an age of 5\,Gyr. The current radii of planets c, d and b are shown as dashed lines, together with the current and final radii for a range of planetary masses for the outer two planets d and b; the core is represented by a black circle, while the envelope is shown as a colored circle around the core. The core size increases with increasing mass. Note that the size scaling of the envelope compared to the core is arbitrary. It is chosen such that one can easily compare present and final planetary radii and immediately see, which configuration can hold on to some fraction of the envelope and remain above the gap. While the outermost planet~b will retain enough envelope to remain well above the radius gap around 1.8\,$R_\oplus$ for all masses considered, we find that for planet~d there is an intermediate mass range between $\sim$7 to 18\,$M_\oplus$, for which planets can retain an envelope and survive above the gap. Lower mass planets will experience enough mass loss to completely lose their envelope, while higher mass planets only have very thin envelopes at present age to match the observed radius due to their larger core size, that their thin envelope is easily lost, in spite of their large mass. Planet c is assumed to be rocky at present and will not change its size significantly. The current radius of planet c matches well with the radii predicted for Earth-like rocky cores in the mass regime of 3-5\,$M_\oplus$.}
\label{fig:R_future}
\end{figure*}

For the inner planet~d, an envelope mass fraction of $\leq$~1.1\% is needed to match the observed radius. The fraction decreases with increasing planet mass, with only 0.02\% needed to reproduce today's radius for a 22\,$M_\oplus$ planet. We show the radius evolution across a range of planetary masses for the middle planet of the K2-198 system in Figure~\ref{fig:R_future} (purple). The results indicate that there is an intermediate mass range, spanning approximately from $\sim$6.6 to 18\,$M_\oplus$, in which planet~d can hold on to some of its atmosphere and continue to reside above the gap with envelope mass fractions between 0.02\% and 0.3\%. Lower mass planets, which do have the largest envelope mass fractions at present age, will lose their remaining envelope by 5\,Gyr. Their low gravitational potential helps boost the mass loss. Interestingly, we see a dichotomy in the planets that lose their envelope for planet~d. In our simulation, the most massive planets also lose their very thin envelopes. While planets with massive cores are generally much better at holding on to their atmospheres, only a very thin envelope is needed to match the observed radius because the bare core itself is already quite large. Even such massive planets cannot hold on to their thin envelopes and will turn into large bare cores by 5\,Gyr.

For the outer planet~c, the results show that across the whole mass range under consideration (7.7 to 24\,$M_\oplus$), and envelope mass fractions in the range of 9-10\%, all planets are able to retain a significant fraction of their envelope by 5\,Gyr and undergo only minor radius evolution. The planets only lose between $\sim$0.1 and 1\% of their total envelope mass, which is negligible in comparison to their large envelopes. This is visualized in red in Figure~\ref{fig:R_future}.

We also conduct all calculations at the younger estimated age of 200\,Myr, and the conservative upper age limit of 500\,Myr. While the mass range for planet~d surviving above the gap is somewhat shifted to lower/higher masses ($\pm 0.4$\,$M_\oplus$ for 200/500\,Myr), a slightly different age does not qualitatively change the finding of an intermediate 'survival' mass range. Planet~b, which experiences less intense evaporation, is not significantly impacted by the starting age of the calculation.

\subsubsection{Ill-constrained past of planets d and b in the absence of planetary masses}
\label{subsec:past}

We further evolve all three planets backwards in time across a range of evolutionary tracks with different spin-down ages (tracks shown in Fig.~\ref{fig:Xray_age}) to investigate how these planets might have started out their lives. Our stopping age for the backwards calculation is 100\,Myr, which is the age limit for which the planetary models are valid. The aim of this backwards calculation or exploration of the planets' past is not to make quantitative predictions, but rather to highlight the degeneracy induced by core mass and envelope mass fraction.

If we assume that the innermost planet c has just evaporated at the present age, our backwards calculation leads to envelope mass fractions at 100\,Myr between 10\% for the lowest activity track, and 20\% for the highest activity track. Due to the high XUV irradiation levels and likely high mass-loss rates at even younger ages, the initial envelope mass fractions were possibly larger. Planet formation models predict initial envelope mass fractions between 0.5 and 9\% \citep{2015LeeChiang, 2020Mordasini, 2019Gupta}, so the innermost planet c would have easily lost an envelope of this mass by the present age. We therefore conclude that planet c has likely been a bare core for many tens of Myr.

\begin{figure}
\includegraphics[width=0.49\textwidth,right]{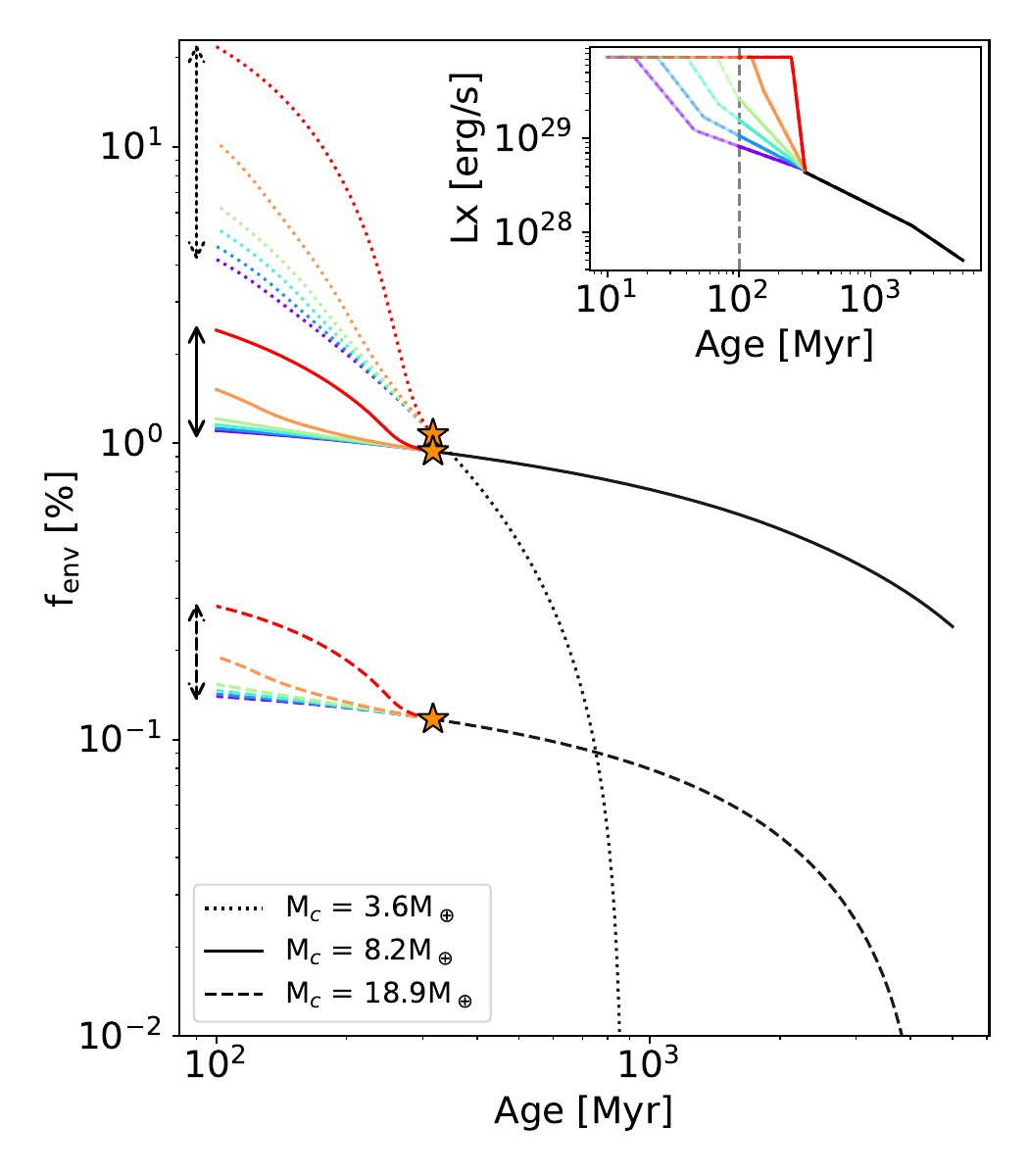}
\includegraphics[width=0.475\textwidth,right]{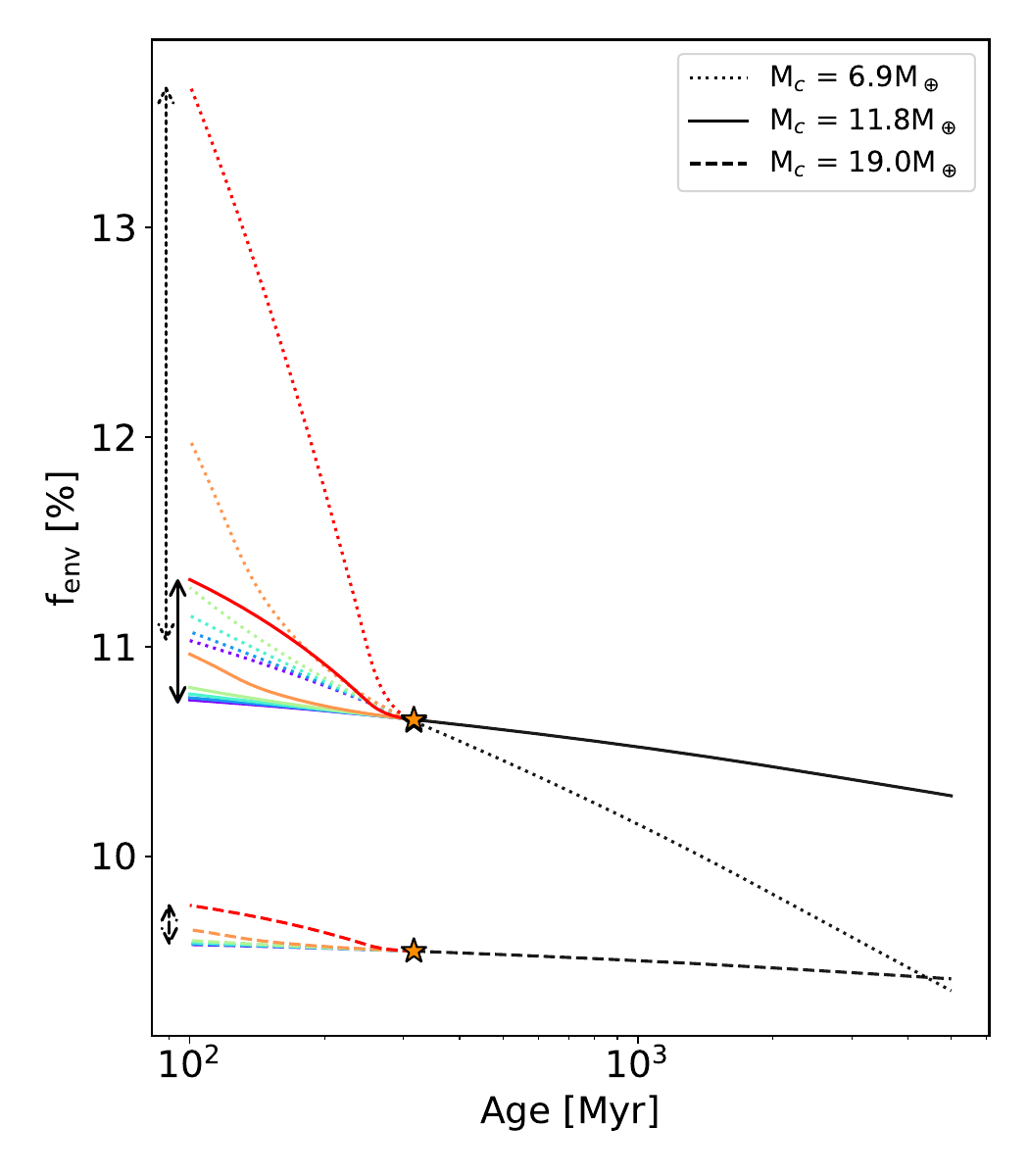}
\caption{Evolution of the envelope mass fraction for planets d and b (upper and lower panel, respectively) as a function of time for three example core masses and six different past activity evolutionary tracks. The different colors represent the evolution along the corresponding activity track, as shown in the embedded $L_X$ vs. age plot in the top right corner. This figure illustrates that without a constraint on the core mass (and to lower extent activity history of the host star), a wide range of possible f$_\mathrm{env}$ and with it radius histories for the two outer planets are possible.}
\label{fig:R_past}
\end{figure}

For planets d and b, such comparison is complicated by the fact that on top of the unknown spin-down age of the star, the planetary masses are unconstrained. If we take three core masses covering the assumed mass ranges, the backwards calculation leads to a range of possible scenarios. This is illustrated for the two outer planets d and b in Figure~\ref{fig:R_past} (top and bottom, respectively). In the figure, a relatively clear trend is visible. Envelope mass fractions, as expected, are higher at younger ages due to the mass loss planets have experienced up to the current age. Lower-mass cores host even higher envelope mass fractions, due to the higher mass-loss rates compared to planets with slightly more massive cores. The range in envelope mass fractions for a particular core mass is the second noticeable result. This difference becomes more pronounced for lower core masses and is a consequence of the chosen stellar activity track, which can range from a low activity, i.e. a short saturation time, to a track with a long saturation time and thus most intense past mass loss. The difference that a stellar activity track makes further decreases with the overall XUV irradiation level -- in the case of planet~b, caused by the larger orbital distance.

For planet~d, the envelope mass fraction at 100 Myr ranges from as little as 0.1-0.3\% to 1-2.5\% to 4-22\% for a planet with an 18.9, 8.2, or 3.6\,$M_\oplus$ core. For planet~b, which is located further away from the host star, the envelope mass fractions range from 9.6-9.8\% to 10.7-11.4\% to 11-14\% for a planet with an 19.0, 11.8, or 7.9\,$M_\oplus$ core -- the spread within each age bin coming from the different possible activity tracks. These values indicate that planet~d, if the core mass is small, might have started out as a large, puffy young planet with a radius of 4-8\,$R_\oplus$ or larger. For heavier core masses, the predicted radii at 100\,Myr are in the sub-Neptune regime ($\sim$2.5-3\,$R_\oplus$). For the outermost planet~b, regardless of the core mass or spin down age, the results indicate a radius in the size regime between Uranus and Saturn at ages around 100\,Myr or younger.

To give a rough idea about how these envelope mass fractions at 100 Myr compare to what is predicted by planet formation models, we provide some numbers. \citet{2015LeeChiang} and \citet{2020Mordasini} (together: LCM) predict primordial envelopes (around the time of disk dissipation, i.e. $\sim$10\,Myr) on the order of 1-2\% for the lowest mass core, while \citet{2019Gupta} (short: GS) predict an envelope as large as 9\% of the total planet mass. For the intermediate mass core, the predicted f$_\mathrm{init}$ is either around 3-4\% from LCM or as high as 12\% from GS. What this comparison tells us is that a low-mass planet with a primordial envelope as thin as predicted by LCM, stands no chance of surviving above the radius gap (with f$_{\mathrm{init}}\sim1\%$) by the current age of the system. If planet~d was a massive planet with a core close to 20\,$M_\oplus$, our backwards calculations produce a 100\,Myr planet with only a very thin envelope with f$_{\mathrm{init}} \leq 0.3\%$. The formation predictions introduced above predict a planet of this core mass to accrete much higher primordial envelopes on the order of 10\% (LCM) or 18\% (GS), suggesting that the true core mass of planet~d is smaller. For planet~b, the LCM formation models predict initial envelopes of 3-4\% (12\%) for the 6.9\,$M_\oplus$ core, 8\% (15\%) for the 11.8\,$M_\oplus$ core, and 15-16\% (18\%) for the massive planet (with GS predictions given in parentheses). If we assume the planet to have hosted envelope mass fractions similar to or higher than the ones we estimate at 100\,Myr, the LCM models suggest a planetary core on the heavier side for planet~b.


We stress that the goal here is not to make any precise quantitative predictions about the past of the planetary system, or try to constrain e.g. the rotational evolution of the host star or core mass of the planets in the system, as has been done previously \citep[e.g.,][]{2020Owen, 2021Bonfanti}. In principle, such comparisons between planet formation model predictions, stellar activity histories and planetary core masses could be used to put constraints on some of these parameters. However, due to all the uncertainties involved in the mass-loss modeling, which includes all the details in modelling the planetary structure itself, any magnetic field effects, the XUV absorption radius of the planet, or the evaporation efficiency at any given age and planet configuration, we refrain from constraining planetary mass or activity evolution and stress the importance of getting a better handle on the planetary masses before conducting more detailed studies.

Overall, our results do show that K2-198~d and b are consistent with a wide range of evaporation histories. We expect planet c to have started out as puffy mini-Neptunes, with planet c likely having lost its primordial atmosphere well before the current age of around 300\,Myr. Planet~d likely started out either as a puffy mini-Neptune or somewhere in the sub-Neptune regime, while planet~b has likely not changed its radius drastically compared to its present-day size. Without measured planetary masses and a more detailed understanding of the rotational spin-down, which includes the factors that influence when a star drops out of the saturated regime, and the timescale for this first rapid spin-down, even multiplanet systems can present an ill-constrained problem in planetary formation and evolution.

\section{Discussion}

With K2-198d residing just above the radius gap and the still relatively high predicted mass-loss rates, K2-198 is an interesting multiplanet system for future detailed hydrodynamic modelling and observations of atmospheric outflow signatures. Although K2-198c has likely lost its primordial envelope due to efficient and intense hydrogen escape at early ages, this planet might host a secondary atmosphere formed by outgassing volatile gases from the magma after the photoevaporation phase \citep{2020Kite,2023Tian}.

Our findings demonstrate how photoevaporative mass loss can lead to the present-day planet parameters starting from a wide range of initial configurations. This is, for one thing, caused by the unconstrained planetary masses, and on the other because stellar spin-down is complex and not fully understood. In particular, the specific spin-down path that stars take from young to old ages may contain time stretches of  rapid versus slow magnetic braking, see for example \citet{Curtis2020, Gruner2020, Dungee2022}.
Our results stress that even for the future evolution of planets in its infant or youth stages, it is crucial to first measure planetary masses, before more constraining studies like those of the multiplanet system around the 400\,Myr-old star HD 63433 (TOI-1726) \citep{2023Damasso}, V1298 Tau \citep{2021Poppenhaeger, 2022Suarez} or K2-102 \citep{2023Brinkman} are possible.

Our mass-loss calculations suggest a dichotomy in planetary mass for planets ending up with and without a remaining primordial atmosphere. Unless the planetary mass is restricted, planets in an intermediate irradiation regime, like K2-198d, could evolve below the gap if their core mass is small enough, or large enough. In the latter case, it is not the low gravitational potential that boosts the mass loss, but instead the low mass of the planetary atmosphere, if it is indeed the large core that makes up for most of the observed planetary radius. For highly-irradiated planets like K2-198c, even for a wide range of planetary core masses such planets cannot hold on to an atmosphere, while for moderately irradiated planets further out, like K2-198b, the planet is stable against complete atmospheric mass loss for a wide range of core masses.

Measuring planetary mass-loss rates across a wide range of planetary parameters, XUV environments and stellar ages will further put constraints on the atmospheric escape model and its modelling inputs, and help distinguish between different mass-loss mechanisms.

\subsubsection{Atmospheric characterization perspectives}

The transiting multiplanet system K2-198, with its young age and relatively high X-ray brightness, is an interesting target for follow-up observations. Suitable options include observing the lower atmosphere layers of the planets with transmission spectra in the infrared with JWST, or trying to observe ongoing mass loss through transmission spectra in the ultraviolet hydrogen Ly-$\alpha$ line or the infrared metastable helium lines (He\,\textsc{i} at 10830\,\AA).

Regarding the characterization of lower-layer atmospheres, based on the observed radii of K2-198d and b and the estimated mass ranges, both planets most likely host volatile envelopes at the present age. We estimate a JWST transmission spectrum metric (TSM; \citet{2018Kempton}) of 13-76 for planet~d and 44-138 for planet~b for the mass ranges under consideration (see Sec.~\ref{sec:masses}). Planet~d, regardless of its estimated mass, has a TSM smaller than the recommended minimum value of 90 for small sub-Neptunes. Depending on the planetary mass, K2-198b, however, could be an interesting target for transmission spectroscopy with JWST. With its radius in the large sub-Neptune regime, the recommended TSM of 90 is exceeded if the planetary mass lies below $\sim 12$\,$M_\oplus$. For planet c, the estimated TSM is 18, which is above the recommended value of 10 for planets with radii smaller than 1.5\,$R_\oplus$, making this planet a favorable target to search for any secondary atmosphere. The emission spectroscopy metric (ESM) for K2-198c, however, is 2.7 and thus below the recommended threshold of 7.5 for terrestrial planets for emission spectroscopy with the James Webb Space Telescope.

Regarding ongoing mass loss, observations of the He\,\textsc{i} lines are promising. It has been suggested that the metastable helium state can be efficiently populated by the stellar spectrum of K-type stars due to the relative fluxes of extreme ultraviolet radiation, which produces triplet He\,\textsc{i} in the ground state of the line, versus the near ultraviolet radiation, which ionizes it \citep{2019Oklopcic}. \citet{2022Poppenhaeger} further suggest that stellar X-ray luminosity and [Fe/O] coronal abundance ratio influence the critical stellar narrow-band EUV emission. They find that young and active stars having [Fe/O] < 1 exhibit lower EUV emission in the 200-504~\AA\, range compared to old and inactive stars with [Fe/O] > 1. K2-198 with its young age and high X-ray luminosity ($\log L_X$\,[\lumi]~$=28.7$) is comparable to the high-activity K-dwarfs in the sample of \citet{2018Wood} that display [Fe/O] ratios lower than unity. This suggests that the steeper EUV-X-ray relation from \citet{2022Poppenhaeger} should be applicable, leading to a lower EUV emission value than for stars with high [Fe/O]. Based on the scaling laws derived in \citet{2022Poppenhaeger}, we estimate that with the current broad-band X-ray flux, planet K2-198d receives a narrow-band EUV flux on the order of 0.5~Wm$^{-1}$, which could translate to relatively high helium absorption scale heights on the order of 60 when compared to known systems with observed helium absorption (see their Fig.~8). For the outermost planet c, we calculate 0.2~Wm$^{-1}$ or roughly 40 helium absorption scale heights. These estimates suggest that, especially K2-198d, although orbiting a young and active star, could be an interesting target for the search of atmospheric mass loss via the metastable helium triplet.

\section{Conclusions}

We use updated eROSITA X-Ray measurements together with photometric and spectroscopic archival data to constrain the age of the multiplanet system K2-198 to 200-500\,Myr. We characterize the present-day high-energy irradiation environment of the three planets, of which the innermost planet is already below the observed exoplanet radius gap in the regime of rocky worlds, while the outer two are mini-Neptunes residing above the gap. With an X-ray luminosity of $\log L_X$\,[\lumi]~$=28.7$, the star has already dropped out of the saturated X-ray regime. We use a model for the stellar activity evolution together with exoplanetary mass loss to estimate the atmospheric evolution of the two outermost planets in the mini-Neptune regime for a range of reasonable planetary masses.

Our calculations indicate that the outermost planet K2-198b will retain a gaseous envelope and survive above the gap over the next several Gyr for planetary masses between $\sim$~7 and 25\,$M_\oplus$. Interestingly, we find that the middle planet K2-198d can only retain an envelope for an intermediate mass range of approximately 7-18\,$M_\oplus$. Lower mass planets experience enough mass loss to become rocky and drop below the radius gap, while higher mass planets, which, at present, would only require a very thin envelope to match the observed radius, are easily stripped despite their larger ability to hold on to their atmospheres. K2-198c, which has likely lost its envelope early on and resides as a bare core below the radius gap, still experiences large amounts of XUV irradiation, and might be an interesting target for investigating the formation of secondary atmospheres.

Due to the lack of measured planetary masses and the unknown stellar activity history, it is challenging to provide constraining predictions to the history of the planetary system. The wide range of compatible evaporation histories for all three planets stresses the importance of getting a better handle on the planetary masses before conducting more detailed studies. Nonetheless, the young K2-198 system is an interesting target to search for ongoing mass loss via metastable helium or possibly Lymann-$\alpha$, which can be expected based on the estimated mass-loss rates, at least for the middle planet K2-198d. Depending on the prospectively measured planetary masses, K2-198d and b could also turn out to be very favorable targets for atmospheric characterization with JWST.


\section*{Acknowledgements}

The authors thank the anonymous referee for providing helpful suggestions on the paper. The authors also thank D. Gruner for compiling a comprehensive list of known parameters for stellar clusters, Clea Schumer for allowing us to use the stellar spectra, and Dus{\'a}n Tub{\'i}n-Arenas for providing eROSITA upper limits. Part of this work was supported by the German \emph{Leibniz-Gemeinschaft}, project number P67-2018.

This work is based on data from eROSITA, the primary instrument aboard SRG, a joint Russian-German science
mission supported by the Russian Space Agency (Roskosmos), in the interests of the Russian Academy of Sciences represented by its Space Research Institute (IKI), and the Deutsches Zentrum f\"ur Luft- und Raumfahrt (DLR). The SRG spacecraft was built by Lavochkin Association (NPOL) and its subcontractors, and is operated by NPOL with support from the Max Planck Institute for Extraterrestrial Physics (MPE). The development and construction of the eROSITA X-ray instrument was led by MPE, with contributions from the Dr. Karl Remeis Observatory Bamberg \& ECAP (FAU Erlangen-N\"urnberg), the University of Hamburg Observatory, the Leibniz Institute for Astrophysics Potsdam (AIP), and the Institute for Astronomy and Astrophysics of the University of T\"ubingen, with the support of DLR and the Max Planck Society. The Argelander Institute for Astronomy of the University of Bonn and the Ludwig Maximilians Universität Munich also participated in the science preparation for eROSITA. The eROSITA data shown here were processed using the eSASS software system developed by the German eROSITA consortium.

This research made use of Lightkurve, a Python package for Kepler and TESS data analysis \citep{2018Lightkurve}, as well as the Python packages \texttt{numpy} \citep{2020numpy}, \texttt{pandas} \citep{mckinney2010data}, \texttt{scipy} \citep{2020scipy}, and \texttt{matplotlib} \citep{hunter2007matplotlib}.
This research has made use of the Exoplanet Follow-up Observation Program (ExoFOP; DOI: 10.26134/ExoFOP5) website, which is operated by the California Institute of Technology, under contract with the National Aeronautics and Space Administration under the Exoplanet Exploration Program, and data from the European Space Agency (ESA) mission {\it Gaia} (\url{https://www.cosmos.esa.int/gaia}), processed by the {\it Gaia}
Data Processing and Analysis Consortium (DPAC,
\url{https://www.cosmos.esa.int/web/gaia/dpac/consortium}). Funding for the DPAC
has been provided by national institutions, in particular the institutions
participating in the {\it Gaia} Multilateral Agreement.


\section*{Data Availability}

This work is based on simulations with the publicly available code "Planetary Photoevaporation Simulator (PLATYPOS)" \citep{2022Ketzer}, which can be accessed on GitHub (\url{https://github.com/lketzer/platypos/}).


\bibliographystyle{mnras}
\bibliography{paper_bib} 

\begin{thebibliography}{}
\makeatletter
\relax
\def\mn@urlcharsother{\let\do\@makeother \do\$\do\&\do\#\do\^\do\_\do\%\do\~}
\def\mn@doi{\begingroup\mn@urlcharsother \@ifnextchar [ {\mn@doi@} {\mn@doi@[]}}
\def\mn@doi@[#1]#2{\def\@tempa{#1}\ifx\@tempa\@empty \href {http://dx.doi.org/#2} {doi:#2}\else \href {http://dx.doi.org/#2} {#1}\fi \endgroup}
\def\mn@eprint#1#2{\mn@eprint@#1:#2::\@nil}
\def\mn@eprint@arXiv#1{\href {http://arxiv.org/abs/#1} {{\tt arXiv:#1}}}
\def\mn@eprint@dblp#1{\href {http://dblp.uni-trier.de/rec/bibtex/#1.xml} {dblp:#1}}
\def\mn@eprint@#1:#2:#3:#4\@nil{\def\@tempa {#1}\def\@tempb {#2}\def\@tempc {#3}\ifx \@tempc \@empty \let \@tempc \@tempb \let \@tempb \@tempa \fi \ifx \@tempb \@empty \def\@tempb {arXiv}\fi \@ifundefined {mn@eprint@\@tempb}{\@tempb:\@tempc}{\expandafter \expandafter \csname mn@eprint@\@tempb\endcsname \expandafter{\@tempc}}}

\bibitem[\protect\citeauthoryear{{Angus} et~al.,}{{Angus} et~al.}{2019}]{2019Angus}
{Angus} R.,  et~al., 2019, \mn@doi [\aj] {10.3847/1538-3881/ab3c53}, \href {https://ui.adsabs.harvard.edu/abs/2019AJ....158..173A} {158, 173}

\bibitem[\protect\citeauthoryear{{Anthony-Twarog}, {Deliyannis}, {Harmer}, {Lee-Brown}, {Steinhauer}, {Sun}  \& {Twarog}}{{Anthony-Twarog} et~al.}{2018}]{2018anthony}
{Anthony-Twarog} B.~J.,  {Deliyannis} C.~P.,  {Harmer} D.,  {Lee-Brown} D.~B.,  {Steinhauer} A.,  {Sun} Q.,   {Twarog} B.~A.,  2018, \mn@doi [\aj] {10.3847/1538-3881/aacb1f}, \href {https://ui.adsabs.harvard.edu/abs/2018AJ....156...37A} {156, 37}

\bibitem[\protect\citeauthoryear{{Asplund}, {Amarsi}  \& {Grevesse}}{{Asplund} et~al.}{2021}]{2021asplund}
{Asplund} M.,  {Amarsi} A.~M.,   {Grevesse} N.,  2021, \mn@doi [\aap] {10.1051/0004-6361/202140445}, \href {https://ui.adsabs.harvard.edu/abs/2021A&A...653A.141A} {653, A141}

\bibitem[\protect\citeauthoryear{{Baraffe}, {Alibert}, {Chabrier}  \& {Benz}}{{Baraffe} et~al.}{2006}]{2006Baraffe}
{Baraffe} I.,  {Alibert} Y.,  {Chabrier} G.,   {Benz} W.,  2006, \mn@doi [\aap] {10.1051/0004-6361:20054040}, \href {https://ui.adsabs.harvard.edu/abs/2006A&A...450.1221B} {450, 1221}

\bibitem[\protect\citeauthoryear{{Baratella} et~al.,}{{Baratella} et~al.}{2020}]{2020baratella_ges}
{Baratella} M.,  et~al., 2020, \mn@doi [\aap] {10.1051/0004-6361/201937055}, \href {https://ui.adsabs.harvard.edu/abs/2020A&A...634A..34B} {634, A34}

\bibitem[\protect\citeauthoryear{{Baratella} et~al.,}{{Baratella} et~al.}{2021}]{2021baratella}
{Baratella} M.,  et~al., 2021, \mn@doi [\aap] {10.1051/0004-6361/202141069}, \href {https://ui.adsabs.harvard.edu/abs/2021A&A...653A..67B} {653, A67}

\bibitem[\protect\citeauthoryear{{Barnes}}{{Barnes}}{2003}]{Barnes2003}
{Barnes} S.~A.,  2003, \mn@doi [\apj] {10.1086/367639}, \href {https://ui.adsabs.harvard.edu/abs/2003ApJ...586..464B} {586, 464}

\bibitem[\protect\citeauthoryear{{Barrag{\'a}n} et~al.,}{{Barrag{\'a}n} et~al.}{2019}]{2019Barragan}
{Barrag{\'a}n} O.,  et~al., 2019, \mn@doi [\mnras] {10.1093/mnras/stz2569}, \href {https://ui.adsabs.harvard.edu/abs/2019MNRAS.490..698B} {490, 698}

\bibitem[\protect\citeauthoryear{{Barrag{\'a}n} et~al.,}{{Barrag{\'a}n} et~al.}{2022}]{2022Barragan}
{Barrag{\'a}n} O.,  et~al., 2022, \mn@doi [\mnras] {10.1093/mnras/stac638}, \href {https://ui.adsabs.harvard.edu/abs/2022MNRAS.514.1606B} {514, 1606}

\bibitem[\protect\citeauthoryear{{Baruteau} et~al.,}{{Baruteau} et~al.}{2014}]{2014Baruteau}
{Baruteau} C.,  et~al., 2014, in {Beuther} H.,  {Klessen} R.~S.,  {Dullemond} C.~P.,   {Henning} T.,  eds, Protostars and Planets VI. pp 667--689 (\mn@eprint {arXiv} {1312.4293}), \mn@doi{10.2458/azu_uapress_9780816531240-ch029}

\bibitem[\protect\citeauthoryear{{Bonfanti}, {Fossati}, {Kubyshkina}  \& {Cubillos}}{{Bonfanti} et~al.}{2021}]{2021Bonfanti}
{Bonfanti} A.,  {Fossati} L.,  {Kubyshkina} D.,   {Cubillos} P.~E.,  2021, \mn@doi [\aap] {10.1051/0004-6361/202142010}, \href {https://ui.adsabs.harvard.edu/abs/2021A&A...656A.157B} {656, A157}

\bibitem[\protect\citeauthoryear{{Bourrier} et~al.,}{{Bourrier} et~al.}{2018}]{2018Bourrier}
{Bourrier} V.,  et~al., 2018, \mn@doi [\aap] {10.1051/0004-6361/201833675}, \href {https://ui.adsabs.harvard.edu/abs/2018A&A...620A.147B} {620, A147}

\bibitem[\protect\citeauthoryear{{Bouvier} et~al.,}{{Bouvier} et~al.}{2018}]{2018bouvier}
{Bouvier} J.,  et~al., 2018, \mn@doi [\aap] {10.1051/0004-6361/201731881}, \href {https://ui.adsabs.harvard.edu/abs/2018A&A...613A..63B} {613, A63}

\bibitem[\protect\citeauthoryear{{Brinkman} et~al.,}{{Brinkman} et~al.}{2023}]{2023Brinkman}
{Brinkman} C.~L.,  et~al., 2023, \mn@doi [\aj] {10.3847/1538-3881/aca64d}, \href {https://ui.adsabs.harvard.edu/abs/2023AJ....165...74B} {165, 74}

\bibitem[\protect\citeauthoryear{{Carolan}, {Vidotto}, {Plavchan}, {Villarreal D'Angelo}  \& {Hazra}}{{Carolan} et~al.}{2020}]{2020Carolan}
{Carolan} S.,  {Vidotto} A.~A.,  {Plavchan} P.,  {Villarreal D'Angelo} C.,   {Hazra} G.,  2020, \mn@doi [\mnras] {10.1093/mnrasl/slaa127}, \href {https://ui.adsabs.harvard.edu/abs/2020MNRAS.498L..53C} {498, L53}

\bibitem[\protect\citeauthoryear{{Casagrande} et~al.,}{{Casagrande} et~al.}{2021}]{2021casagrande}
{Casagrande} L.,  et~al., 2021, \mn@doi [\mnras] {10.1093/mnras/stab2304}, \href {https://ui.adsabs.harvard.edu/abs/2021MNRAS.507.2684C} {507, 2684}

\bibitem[\protect\citeauthoryear{{Castelli} \& {Kurucz}}{{Castelli} \& {Kurucz}}{2003}]{2003castelli}
{Castelli} F.,  {Kurucz} R.~L.,  2003, in {Piskunov} N.,  {Weiss} W.~W.,   {Gray} D.~F.,  eds, Modelling of Stellar Atmospheres. p.~A20 (\mn@eprint {arXiv} {astro-ph/0405087}), \mn@doi{10.48550/arXiv.astro-ph/0405087}

\bibitem[\protect\citeauthoryear{{Chen} \& {Rogers}}{{Chen} \& {Rogers}}{2016}]{2016ChenRogers}
{Chen} H.,  {Rogers} L.~A.,  2016, \mn@doi [\apj] {10.3847/0004-637X/831/2/180}, \href {https://ui.adsabs.harvard.edu/abs/2016ApJ...831..180C} {831, 180}

\bibitem[\protect\citeauthoryear{{Chen} et~al.,}{{Chen} et~al.}{2021}]{2021Chen}
{Chen} D.-C.,  et~al., 2021, \mn@doi [\apj] {10.3847/1538-4357/abd5be}, \href {https://ui.adsabs.harvard.edu/abs/2021ApJ...909..115C} {909, 115}

\bibitem[\protect\citeauthoryear{{Coffaro} et~al.,}{{Coffaro} et~al.}{2020}]{2020Coffaro}
{Coffaro} M.,  et~al., 2020, \mn@doi [\aap] {10.1051/0004-6361/201936479}, \href {https://ui.adsabs.harvard.edu/abs/2020A&A...636A..49C} {636, A49}

\bibitem[\protect\citeauthoryear{{Coffaro}, {Stelzer}  \& {Orlando}}{{Coffaro} et~al.}{2022}]{2022Coffaro}
{Coffaro} M.,  {Stelzer} B.,   {Orlando} S.,  2022, \mn@doi [\aap] {10.1051/0004-6361/202142298}, \href {https://ui.adsabs.harvard.edu/abs/2022A&A...661A..79C} {661, A79}

\bibitem[\protect\citeauthoryear{{Cohen}, {Alvarado-G{\'o}mez}, {Drake}, {Harbach}, {Garraffo}  \& {Fraschetti}}{{Cohen} et~al.}{2022}]{2022Cohen}
{Cohen} O.,  {Alvarado-G{\'o}mez} J.~D.,  {Drake} J.~J.,  {Harbach} L.~M.,  {Garraffo} C.,   {Fraschetti} F.,  2022, \mn@doi [\apj] {10.3847/1538-4357/ac78e4}, \href {https://ui.adsabs.harvard.edu/abs/2022ApJ...934..189C} {934, 189}

\bibitem[\protect\citeauthoryear{{Correia}, {Bourrier}  \& {Delisle}}{{Correia} et~al.}{2020}]{2020Correia}
{Correia} A.~C.~M.,  {Bourrier} V.,   {Delisle} J.~B.,  2020, \mn@doi [\aap] {10.1051/0004-6361/201936967}, \href {https://ui.adsabs.harvard.edu/abs/2020A&A...635A..37C} {635, A37}

\bibitem[\protect\citeauthoryear{{Cummings}, {Deliyannis}, {Maderak}  \& {Steinhauer}}{{Cummings} et~al.}{2017}]{2017cummings}
{Cummings} J.~D.,  {Deliyannis} C.~P.,  {Maderak} R.~M.,   {Steinhauer} A.,  2017, \mn@doi [\aj] {10.3847/1538-3881/aa5b86}, \href {https://ui.adsabs.harvard.edu/abs/2017AJ....153..128C} {153, 128}

\bibitem[\protect\citeauthoryear{{Curtis} et~al.,}{{Curtis} et~al.}{2020a}]{2020Curtis}
{Curtis} J.~L.,  et~al., 2020a, \mn@doi [\apj] {10.3847/1538-4357/abbf58}, \href {https://ui.adsabs.harvard.edu/abs/2020ApJ...904..140C} {904, 140}

\bibitem[\protect\citeauthoryear{{Curtis} et~al.,}{{Curtis} et~al.}{2020b}]{Curtis2020}
{Curtis} J.~L.,  et~al., 2020b, \mn@doi [\apj] {10.3847/1538-4357/abbf58}, \href {https://ui.adsabs.harvard.edu/abs/2020ApJ...904..140C} {904, 140}

\bibitem[\protect\citeauthoryear{{Cutri} et~al.,}{{Cutri} et~al.}{2003}]{2mass}
{Cutri} R.~M.,  et~al., 2003, VizieR Online Data Catalog, \href {https://ui.adsabs.harvard.edu/abs/2003yCat.2246....0C} {p. II/246}

\bibitem[\protect\citeauthoryear{{D'Orazi}, {Magrini}, {Randich}, {Galli}, {Busso}  \& {Sestito}}{{D'Orazi} et~al.}{2009}]{2009dorazi}
{D'Orazi} V.,  {Magrini} L.,  {Randich} S.,  {Galli} D.,  {Busso} M.,   {Sestito} P.,  2009, \mn@doi [\apjl] {10.1088/0004-637X/693/1/L31}, \href {https://ui.adsabs.harvard.edu/abs/2009ApJ...693L..31D} {693, L31}

\bibitem[\protect\citeauthoryear{{D'Orazi}, {Baratella}, {Lugaro}, {Magrini}  \& {Pignatari}}{{D'Orazi} et~al.}{2022}]{2022dorazi}
{D'Orazi} V.,  {Baratella} M.,  {Lugaro} M.,  {Magrini} L.,   {Pignatari} M.,  2022, \mn@doi [Universe] {10.3390/universe8020110}, \href {https://ui.adsabs.harvard.edu/abs/2022Univ....8..110D} {8, 110}

\bibitem[\protect\citeauthoryear{{Damasso} et~al.,}{{Damasso} et~al.}{2023}]{2023Damasso}
{Damasso} M.,  et~al., 2023, \mn@doi [\aap] {10.1051/0004-6361/202245391}, \href {https://ui.adsabs.harvard.edu/abs/2023A&A...672A.126D} {672, A126}

\bibitem[\protect\citeauthoryear{{Davies}, {Adams}, {Armitage}, {Chambers}, {Ford}, {Morbidelli}, {Raymond}  \& {Veras}}{{Davies} et~al.}{2014}]{2014Davies}
{Davies} M.~B.,  {Adams} F.~C.,  {Armitage} P.,  {Chambers} J.,  {Ford} E.,  {Morbidelli} A.,  {Raymond} S.~N.,   {Veras} D.,  2014, in {Beuther} H.,  {Klessen} R.~S.,  {Dullemond} C.~P.,   {Henning} T.,  eds, Protostars and Planets VI. pp 787--808 (\mn@eprint {arXiv} {1311.6816}), \mn@doi{10.2458/azu_uapress_9780816531240-ch034}

\bibitem[\protect\citeauthoryear{{Douglas}, {Curtis}, {Ag{\"u}eros}, {Cargile}, {Brewer}, {Meibom}  \& {Jansen}}{{Douglas} et~al.}{2019}]{2019Douglas}
{Douglas} S.~T.,  {Curtis} J.~L.,  {Ag{\"u}eros} M.~A.,  {Cargile} P.~A.,  {Brewer} J.~M.,  {Meibom} S.,   {Jansen} T.,  2019, \mn@doi [\apj] {10.3847/1538-4357/ab2468}, \href {https://ui.adsabs.harvard.edu/abs/2019ApJ...879..100D} {879, 100}

\bibitem[\protect\citeauthoryear{{Dungee}, {van Saders}, {Gaidos}, {Chun}, {Garc{\'\i}a}, {Magnier}, {Mathur}  \& {Santos}}{{Dungee} et~al.}{2022}]{Dungee2022}
{Dungee} R.,  {van Saders} J.,  {Gaidos} E.,  {Chun} M.,  {Garc{\'\i}a} R.~A.,  {Magnier} E.~A.,  {Mathur} S.,   {Santos} {\^A}. R.~G.,  2022, \mn@doi [\apj] {10.3847/1538-4357/ac90be}, \href {https://ui.adsabs.harvard.edu/abs/2022ApJ...938..118D} {938, 118}

\bibitem[\protect\citeauthoryear{{Ehrenreich} et~al.,}{{Ehrenreich} et~al.}{2015}]{2015Ehrenreich}
{Ehrenreich} D.,  et~al., 2015, \mn@doi [\nat] {10.1038/nature14501}, \href {https://ui.adsabs.harvard.edu/abs/2015Natur.522..459E} {522, 459}

\bibitem[\protect\citeauthoryear{{Fern{\'a}ndez Fern{\'a}ndez} \& {Wheatley}}{{Fern{\'a}ndez Fern{\'a}ndez} \& {Wheatley}}{2022}]{2022Fernandez}
{Fern{\'a}ndez Fern{\'a}ndez} J.,  {Wheatley} P.~J.,  2022, \mn@doi [Astronomische Nachrichten] {10.1002/asna.20210076}, \href {https://ui.adsabs.harvard.edu/abs/2022AN....34310076F} {343, e10076}

\bibitem[\protect\citeauthoryear{{Foreman-Mackey}, {Hogg}, {Lang}  \& {Goodman}}{{Foreman-Mackey} et~al.}{2013}]{2013ForemanMackey}
{Foreman-Mackey} D.,  {Hogg} D.~W.,  {Lang} D.,   {Goodman} J.,  2013, \mn@doi [\pasp] {10.1086/670067}, \href {https://ui.adsabs.harvard.edu/abs/2013PASP..125..306F} {125, 306}

\bibitem[\protect\citeauthoryear{{Fossati} et~al.,}{{Fossati} et~al.}{2017}]{Fossati2017}
{Fossati} L.,  et~al., 2017, \mn@doi [\aap] {10.1051/0004-6361/201629716}, \href {https://ui.adsabs.harvard.edu/abs/2017A&A...598A..90F} {598, A90}

\bibitem[\protect\citeauthoryear{{Foster}, {Poppenhaeger}, {Ilic}  \& {Schwope}}{{Foster} et~al.}{2022a}]{2022Foster}
{Foster} G.,  {Poppenhaeger} K.,  {Ilic} N.,   {Schwope} A.,  2022a, \mn@doi [\aap] {10.1051/0004-6361/202141097}, \href {https://ui.adsabs.harvard.edu/abs/2022A&A...661A..23F} {661, A23}

\bibitem[\protect\citeauthoryear{{Foster}, {Poppenhaeger}, {Ilic}  \& {Schwope}}{{Foster} et~al.}{2022b}]{Foster2022AandA}
{Foster} G.,  {Poppenhaeger} K.,  {Ilic} N.,   {Schwope} A.,  2022b, \mn@doi [\aap] {10.1051/0004-6361/202141097}, \href {https://ui.adsabs.harvard.edu/abs/2022A&A...661A..23F} {661, A23}

\bibitem[\protect\citeauthoryear{{Fritzewski}, {Barnes}, {James}  \& {Strassmeier}}{{Fritzewski} et~al.}{2021}]{2021Fritzewski}
{Fritzewski} D.~J.,  {Barnes} S.~A.,  {James} D.~J.,   {Strassmeier} K.~G.,  2021, \mn@doi [\aap] {10.1051/0004-6361/202140894}, \href {https://ui.adsabs.harvard.edu/abs/2021A&A...652A..60F} {652, A60}

\bibitem[\protect\citeauthoryear{{Fulton} et~al.,}{{Fulton} et~al.}{2017}]{Fulton2017}
{Fulton} B.~J.,  et~al., 2017, \mn@doi [\aj] {10.3847/1538-3881/aa80eb}, \href {https://ui.adsabs.harvard.edu/abs/2017AJ....154..109F} {154, 109}

\bibitem[\protect\citeauthoryear{{Gaia Collaboration} et~al.,}{{Gaia Collaboration} et~al.}{2016}]{2016gaia}
{Gaia Collaboration} et~al., 2016, \mn@doi [\aap] {10.1051/0004-6361/201629272}, \href {https://ui.adsabs.harvard.edu/abs/2016A&A...595A...1G} {595, A1}

\bibitem[\protect\citeauthoryear{{Gaia Collaboration} et~al.,}{{Gaia Collaboration} et~al.}{2022a}]{gaiadr3}
{Gaia Collaboration} et~al., 2022a, \mn@doi [arXiv e-prints] {10.48550/arXiv.2208.00211}, \href {https://ui.adsabs.harvard.edu/abs/2022arXiv220800211G} {p. arXiv:2208.00211}

\bibitem[\protect\citeauthoryear{{Gaia Collaboration} et~al.,}{{Gaia Collaboration} et~al.}{2022b}]{2022Gaia}
{Gaia Collaboration} et~al., 2022b, \mn@doi [arXiv e-prints] {10.48550/arXiv.2208.00211}, \href {https://ui.adsabs.harvard.edu/abs/2022arXiv220800211G} {p. arXiv:2208.00211}

\bibitem[\protect\citeauthoryear{{Gaidos} et~al.,}{{Gaidos} et~al.}{2020}]{2020Gaidos}
{Gaidos} E.,  et~al., 2020, \mn@doi [\mnras] {10.1093/mnras/staa918}, \href {https://ui.adsabs.harvard.edu/abs/2020MNRAS.495..650G} {495, 650}

\bibitem[\protect\citeauthoryear{{Gaidos}, {Claytor}, {Dungee}, {Ali}  \& {Feiden}}{{Gaidos} et~al.}{2023}]{2023Gaidos}
{Gaidos} E.,  {Claytor} Z.,  {Dungee} R.,  {Ali} A.,   {Feiden} G.~A.,  2023, \mn@doi [\mnras] {10.1093/mnras/stad343}, \href {https://ui.adsabs.harvard.edu/abs/2023MNRAS.520.5283G} {520, 5283}

\bibitem[\protect\citeauthoryear{{Garraffo} et~al.,}{{Garraffo} et~al.}{2018}]{2018Garraffo}
{Garraffo} C.,  et~al., 2018, \mn@doi [\apj] {10.3847/1538-4357/aace5d}, \href {https://ui.adsabs.harvard.edu/abs/2018ApJ...862...90G} {862, 90}

\bibitem[\protect\citeauthoryear{{Gillen} et~al.,}{{Gillen} et~al.}{2020}]{2020Gillen}
{Gillen} E.,  et~al., 2020, \mn@doi [\mnras] {10.1093/mnras/stz3251}, \href {https://ui.adsabs.harvard.edu/abs/2020MNRAS.492.1008G} {492, 1008}

\bibitem[\protect\citeauthoryear{{Ginzburg}, {Schlichting}  \& {Sari}}{{Ginzburg} et~al.}{2018}]{2018Ginzburg}
{Ginzburg} S.,  {Schlichting} H.~E.,   {Sari} R.,  2018, \mn@doi [\mnras] {10.1093/mnras/sty290}, \href {https://ui.adsabs.harvard.edu/abs/2018MNRAS.476..759G} {476, 759}

\bibitem[\protect\citeauthoryear{{Gondoin}}{{Gondoin}}{2018}]{2018Gondoin}
{Gondoin} P.,  2018, \mn@doi [\aap] {10.1051/0004-6361/201731541}, \href {https://ui.adsabs.harvard.edu/abs/2018A&A...616A.154G} {616, A154}

\bibitem[\protect\citeauthoryear{{Gruner} \& {Barnes}}{{Gruner} \& {Barnes}}{2020}]{Gruner2020}
{Gruner} D.,  {Barnes} S.~A.,  2020, \mn@doi [\aap] {10.1051/0004-6361/202038984}, \href {https://ui.adsabs.harvard.edu/abs/2020A&A...644A..16G} {644, A16}

\bibitem[\protect\citeauthoryear{{G{\"u}del}}{{G{\"u}del}}{2007}]{Guedel2007}
{G{\"u}del} M.,  2007, \mn@doi [Living Reviews in Solar Physics] {10.12942/lrsp-2007-3}, \href {https://ui.adsabs.harvard.edu/abs/2007LRSP....4....3G} {4, 3}

\bibitem[\protect\citeauthoryear{{Gupta} \& {Schlichting}}{{Gupta} \& {Schlichting}}{2019}]{2019Gupta}
{Gupta} A.,  {Schlichting} H.~E.,  2019, \mn@doi [\mnras] {10.1093/mnras/stz1230}, \href {https://ui.adsabs.harvard.edu/abs/2019MNRAS.487...24G} {487, 24}

\bibitem[\protect\citeauthoryear{Harris et~al.,}{Harris et~al.}{2020}]{2020numpy}
Harris C.~R.,  et~al., 2020, \mn@doi [Nature] {10.1038/s41586-020-2649-2}, 585, 357

\bibitem[\protect\citeauthoryear{{Hedges}, {Saunders}, {Barentsen}, {Coughlin}, {Cardoso}, {Kostov}, {Dotson}  \& {Cody}}{{Hedges} et~al.}{2019}]{2019Hedges}
{Hedges} C.,  {Saunders} N.,  {Barentsen} G.,  {Coughlin} J.~L.,  {Cardoso} J. V. d.~M.,  {Kostov} V.~B.,  {Dotson} J.,   {Cody} A.~M.,  2019, \mn@doi [\apjl] {10.3847/2041-8213/ab2a74}, \href {https://ui.adsabs.harvard.edu/abs/2019ApJ...880L...5H} {880, L5}

\bibitem[\protect\citeauthoryear{{Holmberg}, {Nordstr{\"o}m}  \& {Andersen}}{{Holmberg} et~al.}{2009}]{2009Holmberg}
{Holmberg} J.,  {Nordstr{\"o}m} B.,   {Andersen} J.,  2009, \mn@doi [\aap] {10.1051/0004-6361/200811191}, \href {https://ui.adsabs.harvard.edu/abs/2009A&A...501..941H} {501, 941}

\bibitem[\protect\citeauthoryear{Hunter}{Hunter}{2007}]{hunter2007matplotlib}
Hunter J.~D.,  2007, Computing in science \& engineering, 9, 90

\bibitem[\protect\citeauthoryear{{Iben}}{{Iben}}{1967}]{1967iben}
{Iben} Icko J.,  1967, \mn@doi [\apj] {10.1086/149040}, \href {https://ui.adsabs.harvard.edu/abs/1967ApJ...147..624I} {147, 624}

\bibitem[\protect\citeauthoryear{{Ilic}, {Poppenhaeger}  \& {Hosseini}}{{Ilic} et~al.}{2022}]{2022Ilic}
{Ilic} N.,  {Poppenhaeger} K.,   {Hosseini} S.~M.,  2022, \mn@doi [\mnras] {10.1093/mnras/stac861}, \href {https://ui.adsabs.harvard.edu/abs/2022MNRAS.513.4380I} {513, 4380}

\bibitem[\protect\citeauthoryear{Ilin}{Ilin}{2021}]{Ilin2021}
Ilin E.,  2021, \mn@doi [Journal of Open Source Software] {10.21105/joss.02845}, 6, 2845

\bibitem[\protect\citeauthoryear{{Ilin}, {Schmidt}, {Poppenh{\"a}ger}, {Davenport}, {Kristiansen}  \& {Omohundro}}{{Ilin} et~al.}{2021}]{2021Ilin}
{Ilin} E.,  {Schmidt} S.~J.,  {Poppenh{\"a}ger} K.,  {Davenport} J. R.~A.,  {Kristiansen} M.~H.,   {Omohundro} M.,  2021, \mn@doi [\aap] {10.1051/0004-6361/202039198}, \href {https://ui.adsabs.harvard.edu/abs/2021A&A...645A..42I} {645, A42}

\bibitem[\protect\citeauthoryear{{Jeffries}, {Naylor}, {Mayne}, {Bell}  \& {Littlefair}}{{Jeffries} et~al.}{2013}]{2013jeffries}
{Jeffries} R.~D.,  {Naylor} T.,  {Mayne} N.~J.,  {Bell} C. P.~M.,   {Littlefair} S.~P.,  2013, \mn@doi [\mnras] {10.1093/mnras/stt1180}, \href {https://ui.adsabs.harvard.edu/abs/2013MNRAS.434.2438J} {434, 2438}

\bibitem[\protect\citeauthoryear{{Jenkins} et~al.,}{{Jenkins} et~al.}{2016}]{2016Jenkins}
{Jenkins} J.~M.,  et~al., 2016, in {Chiozzi} G.,  {Guzman} J.~C.,  eds,  Society of Photo-Optical Instrumentation Engineers (SPIE) Conference Series Vol. 9913, Software and Cyberinfrastructure for Astronomy IV. p. 99133E, \mn@doi{10.1117/12.2233418}

\bibitem[\protect\citeauthoryear{{Johnstone}, {Bartel}  \& {G{\"u}del}}{{Johnstone} et~al.}{2021}]{2021Johnstone}
{Johnstone} C.~P.,  {Bartel} M.,   {G{\"u}del} M.,  2021, \mn@doi [\aap] {10.1051/0004-6361/202038407}, \href {https://ui.adsabs.harvard.edu/abs/2021A&A...649A..96J} {649, A96}

\bibitem[\protect\citeauthoryear{{Jontof-Hutter} et~al.,}{{Jontof-Hutter} et~al.}{2016}]{2016JontofHutter}
{Jontof-Hutter} D.,  et~al., 2016, \mn@doi [\apj] {10.3847/0004-637X/820/1/39}, \href {https://ui.adsabs.harvard.edu/abs/2016ApJ...820...39J} {820, 39}

\bibitem[\protect\citeauthoryear{{Kempton} et~al.,}{{Kempton} et~al.}{2018}]{2018Kempton}
{Kempton} E. M.~R.,  et~al., 2018, \mn@doi [\pasp] {10.1088/1538-3873/aadf6f}, \href {https://ui.adsabs.harvard.edu/abs/2018PASP..130k4401K} {130, 114401}

\bibitem[\protect\citeauthoryear{{Ketzer} \& {Poppenhaeger}}{{Ketzer} \& {Poppenhaeger}}{2022}]{2022Ketzer}
{Ketzer} L.,  {Poppenhaeger} K.,  2022, \mn@doi [Astronomische Nachrichten] {10.1002/asna.20210105}, \href {https://ui.adsabs.harvard.edu/abs/2022AN....34310105K} {343, e10105}

\bibitem[\protect\citeauthoryear{{Ketzer} \& {Poppenhaeger}}{{Ketzer} \& {Poppenhaeger}}{2023}]{2023Ketzer}
{Ketzer} L.,  {Poppenhaeger} K.,  2023, \mn@doi [\mnras] {10.1093/mnras/stac2643}, \href {https://ui.adsabs.harvard.edu/abs/2023MNRAS.518.1683K} {518, 1683}

\bibitem[\protect\citeauthoryear{{Kite}, {Fegley}, {Schaefer}  \& {Ford}}{{Kite} et~al.}{2020}]{2020Kite}
{Kite} E.~S.,  {Fegley} Bruce J.,  {Schaefer} L.,   {Ford} E.~B.,  2020, \mn@doi [\apj] {10.3847/1538-4357/ab6ffb}, \href {https://ui.adsabs.harvard.edu/abs/2020ApJ...891..111K} {891, 111}

\bibitem[\protect\citeauthoryear{{Kraft}, {Burrows}  \& {Nousek}}{{Kraft} et~al.}{1991}]{KraftBurrowsNousek1991}
{Kraft} R.~P.,  {Burrows} D.~N.,   {Nousek} J.~A.,  1991, \mn@doi [\apj] {10.1086/170124}, \href {https://ui.adsabs.harvard.edu/abs/1991ApJ...374..344K} {374, 344}

\bibitem[\protect\citeauthoryear{{Kubyshkina} et~al.,}{{Kubyshkina} et~al.}{2019a}]{2019Kubyshkina_b}
{Kubyshkina} D.,  et~al., 2019a, \mn@doi [\aap] {10.1051/0004-6361/201936581}, \href {https://ui.adsabs.harvard.edu/abs/2019A&A...632A..65K} {632, A65}

\bibitem[\protect\citeauthoryear{{Kubyshkina} et~al.,}{{Kubyshkina} et~al.}{2019b}]{2019Kubyshkina_a}
{Kubyshkina} D.,  et~al., 2019b, \mn@doi [\apj] {10.3847/1538-4357/ab1e42}, \href {https://ui.adsabs.harvard.edu/abs/2019ApJ...879...26K} {879, 26}

\bibitem[\protect\citeauthoryear{{Lee} \& {Chiang}}{{Lee} \& {Chiang}}{2015}]{2015LeeChiang}
{Lee} E.~J.,  {Chiang} E.,  2015, \mn@doi [\apj] {10.1088/0004-637X/811/1/41}, \href {https://ui.adsabs.harvard.edu/abs/2015ApJ...811...41L} {811, 41}

\bibitem[\protect\citeauthoryear{{Lee}, {Karalis}  \& {Thorngren}}{{Lee} et~al.}{2022}]{2022Lee}
{Lee} E.~J.,  {Karalis} A.,   {Thorngren} D.~P.,  2022, \mn@doi [\apj] {10.3847/1538-4357/ac9c66}, \href {https://ui.adsabs.harvard.edu/abs/2022ApJ...941..186L} {941, 186}

\bibitem[\protect\citeauthoryear{{Lightkurve Collaboration} et~al.,}{{Lightkurve Collaboration} et~al.}{2018}]{2018Lightkurve}
{Lightkurve Collaboration} et~al., 2018, {Lightkurve: Kepler and TESS time series analysis in Python}, Astrophysics Source Code Library (\mn@eprint {ascl} {1812.013})

\bibitem[\protect\citeauthoryear{{Lind}, {Asplund}  \& {Barklem}}{{Lind} et~al.}{2009}]{2009lind}
{Lind} K.,  {Asplund} M.,   {Barklem} P.~S.,  2009, \mn@doi [\aap] {10.1051/0004-6361/200912221}, \href {https://ui.adsabs.harvard.edu/abs/2009A&A...503..541L} {503, 541}

\bibitem[\protect\citeauthoryear{{Liu}, {Hori}, {Lin}  \& {Asphaug}}{{Liu} et~al.}{2015}]{2015Liu}
{Liu} S.-F.,  {Hori} Y.,  {Lin} D.~N.~C.,   {Asphaug} E.,  2015, \mn@doi [\apj] {10.1088/0004-637X/812/2/164}, \href {https://ui.adsabs.harvard.edu/abs/2015ApJ...812..164L} {812, 164}

\bibitem[\protect\citeauthoryear{{Lopez}}{{Lopez}}{2017}]{2017Lopez}
{Lopez} E.~D.,  2017, \mn@doi [\mnras] {10.1093/mnras/stx1558}, \href {https://ui.adsabs.harvard.edu/abs/2017MNRAS.472..245L} {472, 245}

\bibitem[\protect\citeauthoryear{{Lopez} \& {Fortney}}{{Lopez} \& {Fortney}}{2014}]{LopezFortney2014}
{Lopez} E.~D.,  {Fortney} J.~J.,  2014, \mn@doi [\apj] {10.1088/0004-637X/792/1/1}, \href {https://ui.adsabs.harvard.edu/abs/2014ApJ...792....1L} {792, 1}

\bibitem[\protect\citeauthoryear{{Lopez}, {Fortney}  \& {Miller}}{{Lopez} et~al.}{2012}]{Lopez2012}
{Lopez} E.~D.,  {Fortney} J.~J.,   {Miller} N.,  2012, \mn@doi [\apj] {10.1088/0004-637X/761/1/59}, \href {http://adsabs.harvard.edu/abs/2012ApJ...761...59L} {761, 59}

\bibitem[\protect\citeauthoryear{{Luger}, {Agol}, {Kruse}, {Barnes}, {Becker}, {Foreman-Mackey}  \& {Deming}}{{Luger} et~al.}{2018}]{2018Luger}
{Luger} R.,  {Agol} E.,  {Kruse} E.,  {Barnes} R.,  {Becker} A.,  {Foreman-Mackey} D.,   {Deming} D.,  2018, {EVEREST: Tools for de-trending stellar photometry}, Astrophysics Source Code Library, record ascl:1807.029 (\mn@eprint {ascl} {1807.029})

\bibitem[\protect\citeauthoryear{{Lundkvist} et~al.,}{{Lundkvist} et~al.}{2016}]{2016Lundkvist}
{Lundkvist} M.~S.,  et~al., 2016, \mn@doi [Nature Communications] {10.1038/ncomms11201}, \href {https://ui.adsabs.harvard.edu/abs/2016NatCo...711201L} {7, 11201}

\bibitem[\protect\citeauthoryear{{Magrini} et~al.,}{{Magrini} et~al.}{2018}]{2018magrini}
{Magrini} L.,  et~al., 2018, \mn@doi [\aap] {10.1051/0004-6361/201832841}, \href {https://ui.adsabs.harvard.edu/abs/2018A&A...617A.106M} {617, A106}

\bibitem[\protect\citeauthoryear{{Mansfield} et~al.,}{{Mansfield} et~al.}{2018}]{2018Mansfield}
{Mansfield} M.,  et~al., 2018, \mn@doi [\apjl] {10.3847/2041-8213/aaf166}, \href {https://ui.adsabs.harvard.edu/abs/2018ApJ...868L..34M} {868, L34}

\bibitem[\protect\citeauthoryear{{Marcy} et~al.,}{{Marcy} et~al.}{2014}]{2014Marcy}
{Marcy} G.~W.,  et~al., 2014, \mn@doi [\apjs] {10.1088/0067-0049/210/2/20}, \href {https://ui.adsabs.harvard.edu/abs/2014ApJS..210...20M} {210, 20}

\bibitem[\protect\citeauthoryear{{Matt}, {Brun}, {Baraffe}, {Bouvier}  \& {Chabrier}}{{Matt} et~al.}{2015}]{2015Matt}
{Matt} S.~P.,  {Brun} A.~S.,  {Baraffe} I.,  {Bouvier} J.,   {Chabrier} G.,  2015, \mn@doi [\apjl] {10.1088/2041-8205/799/2/L23}, \href {https://ui.adsabs.harvard.edu/abs/2015ApJ...799L..23M} {799, L23}

\bibitem[\protect\citeauthoryear{{Mayo} et~al.,}{{Mayo} et~al.}{2018}]{2018Mayo}
{Mayo} A.~W.,  et~al., 2018, \mn@doi [\aj] {10.3847/1538-3881/aaadff}, \href {https://ui.adsabs.harvard.edu/abs/2018AJ....155..136M} {155, 136}

\bibitem[\protect\citeauthoryear{{Mazeh}, {Holczer}  \& {Faigler}}{{Mazeh} et~al.}{2016}]{2016Mazeh}
{Mazeh} T.,  {Holczer} T.,   {Faigler} S.,  2016, \mn@doi [\aap] {10.1051/0004-6361/201528065}, \href {https://ui.adsabs.harvard.edu/abs/2016A&A...589A..75M} {589, A75}

\bibitem[\protect\citeauthoryear{McKinney et~al.}{McKinney et~al.}{2010}]{mckinney2010data}
McKinney W.,  et~al., 2010, in Proceedings of the 9th Python in Science Conference. pp 51--56

\bibitem[\protect\citeauthoryear{{Messina}, {Nardiello}, {Desidera}, {Baratella}, {Benatti}, {Biazzo}  \& {D'Orazi}}{{Messina} et~al.}{2022}]{2022Messina}
{Messina} S.,  {Nardiello} D.,  {Desidera} S.,  {Baratella} M.,  {Benatti} S.,  {Biazzo} K.,   {D'Orazi} V.,  2022, \mn@doi [\aap] {10.1051/0004-6361/202142276}, \href {https://ui.adsabs.harvard.edu/abs/2022A&A...657L...3M} {657, L3}

\bibitem[\protect\citeauthoryear{{Mordasini}}{{Mordasini}}{2020}]{2020Mordasini}
{Mordasini} C.,  2020, \mn@doi [\aap] {10.1051/0004-6361/201935541}, \href {https://ui.adsabs.harvard.edu/abs/2020A&A...638A..52M} {638, A52}

\bibitem[\protect\citeauthoryear{{Mordasini}, {Alibert}, {Georgy}, {Dittkrist}, {Klahr}  \& {Henning}}{{Mordasini} et~al.}{2012}]{2012Mordasini}
{Mordasini} C.,  {Alibert} Y.,  {Georgy} C.,  {Dittkrist} K.~M.,  {Klahr} H.,   {Henning} T.,  2012, \mn@doi [\aap] {10.1051/0004-6361/201118464}, \href {https://ui.adsabs.harvard.edu/abs/2012A&A...547A.112M} {547, A112}

\bibitem[\protect\citeauthoryear{{Newton}, {Irwin}, {Charbonneau}, {Berta-Thompson}, {Dittmann}  \& {West}}{{Newton} et~al.}{2016}]{2016Newton}
{Newton} E.~R.,  {Irwin} J.,  {Charbonneau} D.,  {Berta-Thompson} Z.~K.,  {Dittmann} J.~A.,   {West} A.~A.,  2016, \mn@doi [\apj] {10.3847/0004-637X/821/2/93}, \href {https://ui.adsabs.harvard.edu/abs/2016ApJ...821...93N} {821, 93}

\bibitem[\protect\citeauthoryear{{Oklop{\v{c}}i{\'c}}}{{Oklop{\v{c}}i{\'c}}}{2019}]{2019Oklopcic}
{Oklop{\v{c}}i{\'c}} A.,  2019, \mn@doi [\apj] {10.3847/1538-4357/ab2f7f}, \href {https://ui.adsabs.harvard.edu/abs/2019ApJ...881..133O} {881, 133}

\bibitem[\protect\citeauthoryear{{Otegi}, {Bouchy}  \& {Helled}}{{Otegi} et~al.}{2020}]{Otegi2020}
{Otegi} J.~F.,  {Bouchy} F.,   {Helled} R.,  2020, \mn@doi [\aap] {10.1051/0004-6361/201936482}, \href {https://ui.adsabs.harvard.edu/abs/2020A&A...634A..43O} {634, A43}

\bibitem[\protect\citeauthoryear{{Owen} \& {Campos Estrada}}{{Owen} \& {Campos Estrada}}{2020}]{2020Owen}
{Owen} J.~E.,  {Campos Estrada} B.,  2020, \mn@doi [\mnras] {10.1093/mnras/stz3435}, \href {https://ui.adsabs.harvard.edu/abs/2020MNRAS.491.5287O} {491, 5287}

\bibitem[\protect\citeauthoryear{{Owen} \& {Jackson}}{{Owen} \& {Jackson}}{2012}]{2012Owen_Jackson}
{Owen} J.~E.,  {Jackson} A.~P.,  2012, \mn@doi [\mnras] {10.1111/j.1365-2966.2012.21481.x}, \href {https://ui.adsabs.harvard.edu/abs/2012MNRAS.425.2931O} {425, 2931}

\bibitem[\protect\citeauthoryear{{Paegert}, {Stassun}, {Collins}, {Pepper}, {Torres}, {Jenkins}, {Twicken}  \& {Latham}}{{Paegert} et~al.}{2021}]{2021paegert}
{Paegert} M.,  {Stassun} K.~G.,  {Collins} K.~A.,  {Pepper} J.,  {Torres} G.,  {Jenkins} J.,  {Twicken} J.~D.,   {Latham} D.~W.,  2021, \mn@doi [arXiv e-prints] {10.48550/arXiv.2108.04778}, \href {https://ui.adsabs.harvard.edu/abs/2021arXiv210804778P} {p. arXiv:2108.04778}

\bibitem[\protect\citeauthoryear{{Placco}, {Sneden}, {Roederer}, {Lawler}, {Den Hartog}, {Hejazi}, {Maas}  \& {Bernath}}{{Placco} et~al.}{2021}]{2021placco}
{Placco} V.~M.,  {Sneden} C.,  {Roederer} I.~U.,  {Lawler} J.~E.,  {Den Hartog} E.~A.,  {Hejazi} N.,  {Maas} Z.,   {Bernath} P.,  2021, \mn@doi [Research Notes of the American Astronomical Society] {10.3847/2515-5172/abf651}, \href {https://ui.adsabs.harvard.edu/abs/2021RNAAS...5...92P} {5, 92}

\bibitem[\protect\citeauthoryear{{Poppenhaeger}}{{Poppenhaeger}}{2022}]{2022Poppenhaeger}
{Poppenhaeger} K.,  2022, \mn@doi [\mnras] {10.1093/mnras/stac507}, \href {https://ui.adsabs.harvard.edu/abs/2022MNRAS.512.1751P} {512, 1751}

\bibitem[\protect\citeauthoryear{{Poppenhaeger} \& {Schmitt}}{{Poppenhaeger} \& {Schmitt}}{2011}]{2011Poppenhaeger}
{Poppenhaeger} K.,  {Schmitt} J.~H.~M.~M.,  2011, \mn@doi [Astronomische Nachrichten] {10.1002/asna.201111615}, \href {https://ui.adsabs.harvard.edu/abs/2011AN....332.1052P} {332, 1052}

\bibitem[\protect\citeauthoryear{{Poppenhaeger}, {Ketzer}  \& {Mallonn}}{{Poppenhaeger} et~al.}{2021}]{2021Poppenhaeger}
{Poppenhaeger} K.,  {Ketzer} L.,   {Mallonn} M.,  2021, \mn@doi [\mnras] {10.1093/mnras/staa1462}, \href {https://ui.adsabs.harvard.edu/abs/2021MNRAS.500.4560P} {500, 4560}

\bibitem[\protect\citeauthoryear{{Predehl} et~al.,}{{Predehl} et~al.}{2021}]{Predehl2021}
{Predehl} P.,  et~al., 2021, \mn@doi [\aap] {10.1051/0004-6361/202039313}, \href {https://ui.adsabs.harvard.edu/abs/2021A&A...647A...1P} {647, A1}

\bibitem[\protect\citeauthoryear{{Reddy} \& {Lambert}}{{Reddy} \& {Lambert}}{2015}]{2015reddy}
{Reddy} A. B.~S.,  {Lambert} D.~L.,  2015, \mn@doi [\mnras] {10.1093/mnras/stv1876}, \href {https://ui.adsabs.harvard.edu/abs/2015MNRAS.454.1976R} {454, 1976}

\bibitem[\protect\citeauthoryear{{Reddy} \& {Lambert}}{{Reddy} \& {Lambert}}{2017}]{2017reddy}
{Reddy} A. B.~S.,  {Lambert} D.~L.,  2017, \mn@doi [\apj] {10.3847/1538-4357/aa81d6}, \href {https://ui.adsabs.harvard.edu/abs/2017ApJ...845..151R} {845, 151}

\bibitem[\protect\citeauthoryear{{Rogers}}{{Rogers}}{2015}]{2015Rogers}
{Rogers} L.~A.,  2015, \mn@doi [\apj] {10.1088/0004-637X/801/1/41}, \href {https://ui.adsabs.harvard.edu/abs/2015ApJ...801...41R} {801, 41}

\bibitem[\protect\citeauthoryear{{Romano} et~al.,}{{Romano} et~al.}{2021}]{2021romano}
{Romano} D.,  et~al., 2021, \mn@doi [\aap] {10.1051/0004-6361/202141340}, \href {https://ui.adsabs.harvard.edu/abs/2021A&A...653A..72R} {653, A72}

\bibitem[\protect\citeauthoryear{{Sanchis-Ojeda} et~al.,}{{Sanchis-Ojeda} et~al.}{2013}]{2013SanchisOjeda}
{Sanchis-Ojeda} R.,  et~al., 2013, \mn@doi [\apj] {10.1088/0004-637X/775/1/54}, \href {https://ui.adsabs.harvard.edu/abs/2013ApJ...775...54S} {775, 54}

\bibitem[\protect\citeauthoryear{{Sanz-Forcada}, {Stelzer}, {Coffaro}, {Raetz}  \& {Alvarado-G{\'o}mez}}{{Sanz-Forcada} et~al.}{2019}]{2019SanzForcada}
{Sanz-Forcada} J.,  {Stelzer} B.,  {Coffaro} M.,  {Raetz} S.,   {Alvarado-G{\'o}mez} J.~D.,  2019, \mn@doi [\aap] {10.1051/0004-6361/201935703}, \href {https://ui.adsabs.harvard.edu/abs/2019A&A...631A..45S} {631, A45}

\bibitem[\protect\citeauthoryear{{Shibayama} et~al.,}{{Shibayama} et~al.}{2013}]{2013Shibayama}
{Shibayama} T.,  et~al., 2013, \mn@doi [\apjs] {10.1088/0067-0049/209/1/5}, \href {https://ui.adsabs.harvard.edu/abs/2013ApJS..209....5S} {209, 5}

\bibitem[\protect\citeauthoryear{{Shkolnik} \& {Llama}}{{Shkolnik} \& {Llama}}{2018}]{2018Shkolnik}
{Shkolnik} E.~L.,  {Llama} J.,  2018, in {Deeg} H.~J.,  {Belmonte} J.~A.,  eds, , Handbook of Exoplanets.
p.~20, \mn@doi{10.1007/978-3-319-55333-7_20}

\bibitem[\protect\citeauthoryear{{Sneden}}{{Sneden}}{1973}]{1973sneden}
{Sneden} C.~A.,  1973, PhD thesis, University of Texas, Austin

\bibitem[\protect\citeauthoryear{{Spake} et~al.,}{{Spake} et~al.}{2018}]{2018Spake}
{Spake} J.~J.,  et~al., 2018, \mn@doi [\nat] {10.1038/s41586-018-0067-5}, \href {https://ui.adsabs.harvard.edu/abs/2018Natur.557...68S} {557, 68}

\bibitem[\protect\citeauthoryear{{Stauffer} et~al.,}{{Stauffer} et~al.}{2016}]{2016Stauffer}
{Stauffer} J.,  et~al., 2016, \mn@doi [\aj] {10.3847/0004-6256/151/3/60}, \href {https://ui.adsabs.harvard.edu/abs/2016AJ....151...60S} {151, 60}

\bibitem[\protect\citeauthoryear{{Str{\"o}mberg}}{{Str{\"o}mberg}}{1946}]{1946Stromberg}
{Str{\"o}mberg} G.,  1946, \mn@doi [\apj] {10.1086/144830}, \href {https://ui.adsabs.harvard.edu/abs/1946ApJ...104...12S} {104, 12}

\bibitem[\protect\citeauthoryear{{Strugarek}, {Bolmont}, {Mathis}, {Brun}, {R{\'e}ville}, {Gallet}  \& {Charbonnel}}{{Strugarek} et~al.}{2017}]{2017Strugarek}
{Strugarek} A.,  {Bolmont} E.,  {Mathis} S.,  {Brun} A.~S.,  {R{\'e}ville} V.,  {Gallet} F.,   {Charbonnel} C.,  2017, \mn@doi [\apjl] {10.3847/2041-8213/aa8d70}, \href {https://ui.adsabs.harvard.edu/abs/2017ApJ...847L..16S} {847, L16}

\bibitem[\protect\citeauthoryear{{Stumpe} et~al.,}{{Stumpe} et~al.}{2012}]{2012Stumpe}
{Stumpe} M.~C.,  et~al., 2012, \mn@doi [\pasp] {10.1086/667698}, \href {https://ui.adsabs.harvard.edu/abs/2012PASP..124..985S} {124, 985}

\bibitem[\protect\citeauthoryear{{Su{\'a}rez Mascare{\~n}o} et~al.,}{{Su{\'a}rez Mascare{\~n}o} et~al.}{2021}]{2022Suarez}
{Su{\'a}rez Mascare{\~n}o} A.,  et~al., 2021, \mn@doi [Nature Astronomy] {10.1038/s41550-021-01533-7}, \href {https://ui.adsabs.harvard.edu/abs/2022NatAs...6..232S} {6, 232}

\bibitem[\protect\citeauthoryear{{Sunyaev} et~al.,}{{Sunyaev} et~al.}{2021}]{Sunyaev2021}
{Sunyaev} R.,  et~al., 2021, \mn@doi [\aap] {10.1051/0004-6361/202141179}, \href {https://ui.adsabs.harvard.edu/abs/2021A&A...656A.132S} {656, A132}

\bibitem[\protect\citeauthoryear{{Tian} \& {Heng}}{{Tian} \& {Heng}}{2023}]{2023Tian}
{Tian} M.,  {Heng} K.,  2023, \mn@doi [arXiv e-prints] {10.48550/arXiv.2301.10217}, \href {https://ui.adsabs.harvard.edu/abs/2023arXiv230110217T} {p. arXiv:2301.10217}

\bibitem[\protect\citeauthoryear{{Tu}, {Johnstone}, {G{\"u}del}  \& {Lammer}}{{Tu} et~al.}{2015}]{2015Tu}
{Tu} L.,  {Johnstone} C.~P.,  {G{\"u}del} M.,   {Lammer} H.,  2015, \mn@doi [\aap] {10.1051/0004-6361/201526146}, \href {https://ui.adsabs.harvard.edu/abs/2015A&A...577L...3T} {577, L3}

\bibitem[\protect\citeauthoryear{{Van Eylen}, {Agentoft}, {Lundkvist}, {Kjeldsen}, {Owen}, {Fulton}, {Petigura}  \& {Snellen}}{{Van Eylen} et~al.}{2018}]{VanEylen2018b}
{Van Eylen} V.,  {Agentoft} C.,  {Lundkvist} M.~S.,  {Kjeldsen} H.,  {Owen} J.~E.,  {Fulton} B.~J.,  {Petigura} E.,   {Snellen} I.,  2018, \mn@doi [\mnras] {10.1093/mnras/sty1783}, \href {https://ui.adsabs.harvard.edu/abs/2018MNRAS.479.4786V} {479, 4786}

\bibitem[\protect\citeauthoryear{{Vanderburg} \& {Johnson}}{{Vanderburg} \& {Johnson}}{2014}]{2014Vanderburg}
{Vanderburg} A.,  {Johnson} J.~A.,  2014, \mn@doi [\pasp] {10.1086/678764}, \href {https://ui.adsabs.harvard.edu/abs/2014PASP..126..948V} {126, 948}

\bibitem[\protect\citeauthoryear{{Venturini}, {Guilera}, {Haldemann}, {Ronco}  \& {Mordasini}}{{Venturini} et~al.}{2020}]{2020Venturini}
{Venturini} J.,  {Guilera} O.~M.,  {Haldemann} J.,  {Ronco} M.~P.,   {Mordasini} C.,  2020, \mn@doi [\aap] {10.1051/0004-6361/202039141}, \href {https://ui.adsabs.harvard.edu/abs/2020A&A...643L...1V} {643, L1}

\bibitem[\protect\citeauthoryear{Virtanen et~al.,}{Virtanen et~al.}{2020}]{2020scipy}
Virtanen P.,  et~al., 2020, \mn@doi [Nature Methods] {10.1038/s41592-019-0686-2}, \href {https://rdcu.be/b08Wh} {17, 261}

\bibitem[\protect\citeauthoryear{{Vissapragada} et~al.,}{{Vissapragada} et~al.}{2021}]{2021Vissapragada}
{Vissapragada} S.,  et~al., 2021, \mn@doi [\aj] {10.3847/1538-3881/ac1bb0}, \href {https://ui.adsabs.harvard.edu/abs/2021AJ....162..222V} {162, 222}

\bibitem[\protect\citeauthoryear{{Watson}, {Donahue}  \& {Walker}}{{Watson} et~al.}{1981}]{Watson1981}
{Watson} A.~J.,  {Donahue} T.~M.,   {Walker} J.~C.~G.,  1981, \mn@doi [\icarus] {10.1016/0019-1035(81)90101-9}, \href {https://ui.adsabs.harvard.edu/abs/1981Icar...48..150W} {48, 150}

\bibitem[\protect\citeauthoryear{{Wood}, {Laming}, {Warren}  \& {Poppenhaeger}}{{Wood} et~al.}{2018}]{2018Wood}
{Wood} B.~E.,  {Laming} J.~M.,  {Warren} H.~P.,   {Poppenhaeger} K.,  2018, \mn@doi [\apj] {10.3847/1538-4357/aaccf6}, \href {https://ui.adsabs.harvard.edu/abs/2018ApJ...862...66W} {862, 66}

\bibitem[\protect\citeauthoryear{{Wright}, {Drake}, {Mamajek}  \& {Henry}}{{Wright} et~al.}{2011a}]{Wright2011}
{Wright} N.~J.,  {Drake} J.~J.,  {Mamajek} E.~E.,   {Henry} G.~W.,  2011a, \mn@doi [\apj] {10.1088/0004-637X/743/1/48}, \href {http://adsabs.harvard.edu/abs/2011ApJ...743...48W} {743, 48}

\bibitem[\protect\citeauthoryear{{Wright}, {Drake}, {Mamajek}  \& {Henry}}{{Wright} et~al.}{2011b}]{2011Wright}
{Wright} N.~J.,  {Drake} J.~J.,  {Mamajek} E.~E.,   {Henry} G.~W.,  2011b, \mn@doi [\apj] {10.1088/0004-637X/743/1/48}, \href {https://ui.adsabs.harvard.edu/abs/2011ApJ...743...48W} {743, 48}

\bibitem[\protect\citeauthoryear{{Wright}, {Newton}, {Williams}, {Drake}  \& {Yadav}}{{Wright} et~al.}{2018}]{2018Wright}
{Wright} N.~J.,  {Newton} E.~R.,  {Williams} P. K.~G.,  {Drake} J.~J.,   {Yadav} R.~K.,  2018, \mn@doi [\mnras] {10.1093/mnras/sty1670}, \href {https://ui.adsabs.harvard.edu/abs/2018MNRAS.479.2351W} {479, 2351}

\bibitem[\protect\citeauthoryear{{Wyatt}, {Kral}  \& {Sinclair}}{{Wyatt} et~al.}{2020}]{2020Wyatt}
{Wyatt} M.~C.,  {Kral} Q.,   {Sinclair} C.~A.,  2020, \mn@doi [\mnras] {10.1093/mnras/stz3052}, \href {https://ui.adsabs.harvard.edu/abs/2020MNRAS.491..782W} {491, 782}

\bibitem[\protect\citeauthoryear{{Zeng} et~al.,}{{Zeng} et~al.}{2019}]{2019Zeng}
{Zeng} L.,  et~al., 2019, \mn@doi [Proceedings of the National Academy of Science] {10.1073/pnas.1812905116}, \href {https://ui.adsabs.harvard.edu/abs/2019PNAS..116.9723Z} {116, 9723}

\bibitem[\protect\citeauthoryear{{Zhang} et~al.,}{{Zhang} et~al.}{2022}]{2022Zhang}
{Zhang} M.,  et~al., 2022, \mn@doi [\aj] {10.3847/1538-3881/ac3f3b}, \href {https://ui.adsabs.harvard.edu/abs/2022AJ....163...68Z} {163, 68}

\bibitem[\protect\citeauthoryear{{Zhang}, {Knutson}, {Dai}, {Wang}, {Ricker}, {Schwarz}, {Mann}  \& {Collins}}{{Zhang} et~al.}{2023}]{2023Zhang}
{Zhang} M.,  {Knutson} H.~A.,  {Dai} F.,  {Wang} L.,  {Ricker} G.~R.,  {Schwarz} R.~P.,  {Mann} C.,   {Collins} K.,  2023, \mn@doi [\aj] {10.3847/1538-3881/aca75b}, \href {https://ui.adsabs.harvard.edu/abs/2023AJ....165...62Z} {165, 62}

\makeatother
\end{thebibliography}


\bsp	
\label{lastpage}
\end{document}